\documentclass[aps,pre,twocolumn,longbibliography,superscriptaddress]{revtex4-2}
\usepackage{graphicx,amsmath,hyperref,xcolor, float}
\begin{document}
\title{Dynamics of self-organization in dense persistent active matter}
\author{Atharva Shukla}
\affiliation{%
 Waltham, Massachusetts 02453, USA}
\author{Chandan Dasgupta}
\affiliation{%
Department of Physics, Indian Institute of Science, Bangalore 560012, India}
\affiliation{%
International Centre for Theoretical Sciences, TIFR, Bangalore 560089, India}
\date{\today}
\begin{abstract}
    We consider a two-dimensional athermal binary mixture of Lennard-Jones particles with persistent random active forces. The liquid phase of this system for active forces exceeding a threshold value exhibits self-organization with long-range spatial correlations of particle velocities and active forces. We study by simulations the development of these correlations from a random initial state.  Several characteristics of the growth of correlations are measured and compared with those of phase-ordering kinetics of equilibrium systems after a quench from a disordered state. The motion of the particles in the long-time steady state is found to be dominated by two streams that flow in opposite directions.
\end{abstract}
\maketitle
\section{Introduction}
\label{sec:introduction} 
Glassy dynamics and jamming in dense active systems of self-propelled particles have received a great deal of attention in recent years~\cite{janssen2019,chaudhuri2021}. This interest is driven by a large number of experimental observations.
In several biological systems, such as bacterial cytoplasm~\cite{parry2014,nishizawa2017}, cell nucleus~\cite{hameed2012}, epithelial sheets of cells ~\cite{angelini2011,garcia2015,park2015,henkes2020}, and bacterial assemblies~\cite{lama2024}, self-propulsion or activity is found to fluidize a glassy state that exhibits characteristic glassy features in the absence of activity. Recent experiments on dense systems of Janus colloids~\cite{klongvessa2019a,klongvessa2019b} and vibrated granular systems~\cite{arora2022} have provided a lot of information about how activity affects glassy dynamics and jamming. To develop a theoretical understanding of these non-equilibrium phenomena, the effects of activity in several model glass-forming liquids have been studied~\cite{henkes2011,ni2013,berthier2014,berthier2017,mandal2016,mandal2017} using molecular dynamics and Brownian dynamics simulations.  The activity in these systems is characterized by two parameters: the magnitude $f$ of the active force and its persistence time $\tau_p$. If the persistence time is short, then the observed behavior is similar to that near the usual glass transition in passive systems. 
Some of the effects of activity on the glass transition for small and medium values of $\tau_p$ can be understood from a generalization of the Random First-Order Transition (RFOT) theory of the glass transition to active systems~\cite{nandi2018,mandal2022}. For long persistence times, the approach to dynamic arrest at low propulsion force goes through a phase characterized by intermittency~\cite{mandal2020,keta2022,keta2023}. The intermittency is a consequence of long periods of jamming followed by bursts of plastic yielding. 
In the limit of infinite persistence time in which the active forces on the particles do not change in time, the homogeneous liquid state obtained for an athermal system at high values of the active force evolves to a force-balanced jammed state when the self-propulsion force decreases below a threshold value~\cite{mandal2020,liao2018,villarroel2021,yang2022}.

In this paper, we report the results of a numerical study of the dynamics of a two-dimensional active system in the liquid state obtained at large values of the self-propulsion force in the limit of infinite persistence time. In our system, the active forces on all particles have the same magnitude, but randomly chosen directions that do not change in time. The liquid state of this system exhibits an interesting self-organization~\cite{suman2025}: particle velocities and self-propulsion forces develop long-range spatial correlations in the steady state, and length scales extracted from these correlations are found to increase with system size without showing any sign of saturation. This behavior is consistent with previous studies~\cite{henkes2020,yang2022,szamel2021,caprini2020a,caprini2020b,kuroda2023} showing that the length scale associated with spatial correlation of the velocity field increases with $\tau_p$ if $\tau_p$ is large. In this paper, we present numerical results for the development of this self-organized state from an initial state in which such correlations are not present. We note that the development of spatial correlations in the directions of the active forces is more unusual than the development of similar correlations in the velocity field. Particle velocities can align through collisions, but alignment of the active forces requires physical motion of particles with similar directions of active forces, some of which may be separated by large distances in the initial state, to the same region of space. We have studied how the length scales associated with correlations of velocities and active forces increase with time as the system evolves to the steady state from a random initial state. This process is analogous to the development of order~\cite{bray2002} in a system after a quench from an initial disordered state to a point inside the ordered region of the phase diagram. We have also studied the characteristics of the velocity field in the steady state and found evidence suggesting that the motion is dominated by two streams of particles moving in opposite directions.

The main results obtained from our study are summarized below. We find that the growth of spatial correlations of particle velocities and active forces exhibits features similar to those found in the growth of ordered domains~\cite{bray2002} after a quench from a disordered phase in systems that exhibit an order-disorder transition in equilibrium. In particular, the development of correlations exhibits self-similarity in time, with a length scale that initially increases with time and saturates at long time. The fields that get correlated in space in our system (the velocity and the active force) are two-dimensional vectors in the two-dimensional system we consider. The sums of these two-dimensional vectors are fixed at zero in our simulations. So, the symmetry of the ordering field in our system is that of the conserved XY model~\cite{bray1989,bray1990,mondello1993,siegert1993} in two dimensions. Our results for the temporal growth of the correlation lengths in our system are different from those found in analytic~\cite{bray1989,bray1990} and numerical~\cite{mondello1993,siegert1993} studies of the conserved XY model in two dimensions. The growth of velocity correlations is consistent with Porod's law~\cite{bray2002,porod83}, which indicates smooth domain boundaries. However, the growth of correlations of the active force shows a clear deviation from Porod's law, suggesting that the boundaries of the domains in which the active forces are nearly parallel with one another are rough. The distribution of the angles of the velocities of the particles in the steady state is not isotropic - it exhibits two peaks in directions separated by 180 degrees. This anisotropy does not decrease as the system size is increased. This observation implies that the motion of the particles in the steady state is dominated by two streams that flow in opposite directions. This suggests a similarity with lane formation~\cite{dzubiella2002} in a mixture of particles with opposite charge in the presence of an external electric field.

The remainder of the paper is organized as follows. The model we have studied and the method of simulation are described in section~\ref{model}. The main results are described in detail in section~\ref{results}. The conclusions derived from the results are summarized in section~\ref{conclusions}.

\section{Model and Simulation Details}
\label{model}

We study a binary system of N  particles~\cite{mandal2020,suman2025} in a 2D square box of size L $\times$ L. The binary mixture consists of two types of particles, A and B, in the ratio 65:35. These particles interact via a Lennard-Jones potential
\begin{equation}
    V_{ij}(r) = 4\epsilon_{\alpha\beta}\left[\left(\frac{\sigma_{\alpha\beta}}{r}\right)^{12} -\left(\frac{\sigma_{\alpha\beta}}{r}\right)^6\right]
\end{equation}
where $\alpha,\beta \in {A,B}$, $r=|\textbf{r}_i - \textbf{r}_j|$ is the distance between the $i$ and $j$ particles. The values of the parameters that define the range and the strength of the interactions are $\sigma_{AB} = 0.8$, $\sigma_{BB} = 0.88$, $\epsilon_{AB} = 1.5$ and $\epsilon_{BB} = 0.5$ in units where $\sigma_{AA} = 1$ and $\epsilon_{AA} = 1$. The interactions are truncated at $r_c = 2.5\sigma_{\alpha\beta}$. All particles have the same mass $(m=1)$. The average particle density $\rho$ is fixed at 1.2. These model parameters are taken from the Kob-Anderson model~\cite{Brüning_2009} that shows dynamical arrest at low temperatures.

The particles in our system are driven by self-propulsion forces ${\textbf{f}_{a,i}}$ =$f\textbf{n}_i$, where $\textbf{n}_i = (\cos\theta_i, \sin\theta_i)$. The directions $\{\textbf{n}_i\}$ of the active forces are chosen randomly and uniformly between $0$ and $2\pi$ with $\sum_i \textbf{n}_i =0 $ and they remain fixed throughout the simulation. The equation of motion of the particles in our athermal system are given by
\begin{equation}
    m{\ddot{{\textbf{r}}_i}} = -\gamma{\dot{{\bf{r}}_i}}+\sum_{j\neq i}^{\text{N}}{\textbf{f}}_{ij}+f{{\textbf{n}}_i}.
\end{equation}
Here, $\textbf{f}_{ij}$ is the inter-particle force modelled by the Lennard-Jones interactions. The units chosen to non-dimensionalise the equation of motion are as follows: $M = m$, $L=\sigma_{AA}$, $E = \epsilon_{AA}$, $\tau = \sigma_{AA}\sqrt{\frac{m}{\epsilon_{AA}}}$.
The magnitude of the self-propulsion force is set at $f = 3$ for which the system is in the liquid state. Since this value is substantially larger than that at which a liquid to jammed solid transition takes place~\cite{mandal2020}, the observed behaviour is not affected by proximity to the jamming transition.
 The friction coefficient $\gamma = 1$ and the integration timestep $\Delta t \in$ 0.001,0.0006 depending on the system size. The position and velocity update rules are as follows~\cite{mandal2021}: 
\begin{equation}
\label{eq:positionupdate}
    {\bf{r}}_i(t+\Delta t) = {\bf{r}}_i(t)+c_1{\bf{v}}_i(t)+c_2\left({\bf{f}}_{a,i}+\sum_{i \neq j}^{N} {\bf{f}}_{ij}(t)\right)
\end{equation}
\begin{equation}
    \label{eq:velocityupdate}
    {\bf{v}}_i(t+\Delta t) = \tau{\bf{v}}_i(t) + \frac{1}{\gamma}(1-\tau)\left({\bf{f}}_{a,i}+\sum_{i \neq j}^{N} {\bf{f}}_{ij}(t)\right)
\end{equation}
where $\tau=e^{-{\frac{\gamma\Delta t}{m}}}$, $c_1=\frac{m}{\gamma}(1-\tau)$ and $c_2=\frac{m}{\gamma^2}\left(\gamma\frac{\Delta t}{m}-1+\tau\right)$. A derivation of these update rules is provided in Appendix I.\\

Starting from a crystalline lattice of the particles with randomly assigned velocities, we simulate the passive thermal dynamics of the system at temperature $T=0.25$ in reduced units. When the kinetic energy of the particles stabilizes, active forces are added to each particle, adhering to the conditions associated with them described earlier. This system evolves as specified by Eqs.\eqref{eq:positionupdate} and \eqref{eq:velocityupdate}.
The evolution of the system is monitored until it reaches a steady state. We assume that a steady state has been reached when the equal time correlation functions described below and the pair distribution functions that describe the structure of the liquid become independent of time. 

\section{Methods and Results}
\label{results}

In this section, we outline the methods used in our analysis of the simulation data and describe the main results. Most of the results presented below are for a 2D system of 38880 particles in a square box of dimension $L = 180$. Systems with 1080 particles ($L = 30$) and 9720 particles ($L = 90$) were also simulated to check finite-size effects. The reported results are averaged over 78 different runs. Each run has different directions of active forces. The initial lattice positions of the particles are randomly chosen from 5 different configurations.
\subsection{Visualizing the growth of velocity and force domains}
To visualize how the system evolves, we show snapshots of the particle system in which the color of the dot representing a particle corresponds to the angle made by the velocity or active force of the particle with the x-axis. The regions where there is a congregation of particles moving in similar directions or having similar directions of active forces can easily be seen. 

\begin{figure*}
\includegraphics[width=2\columnwidth]{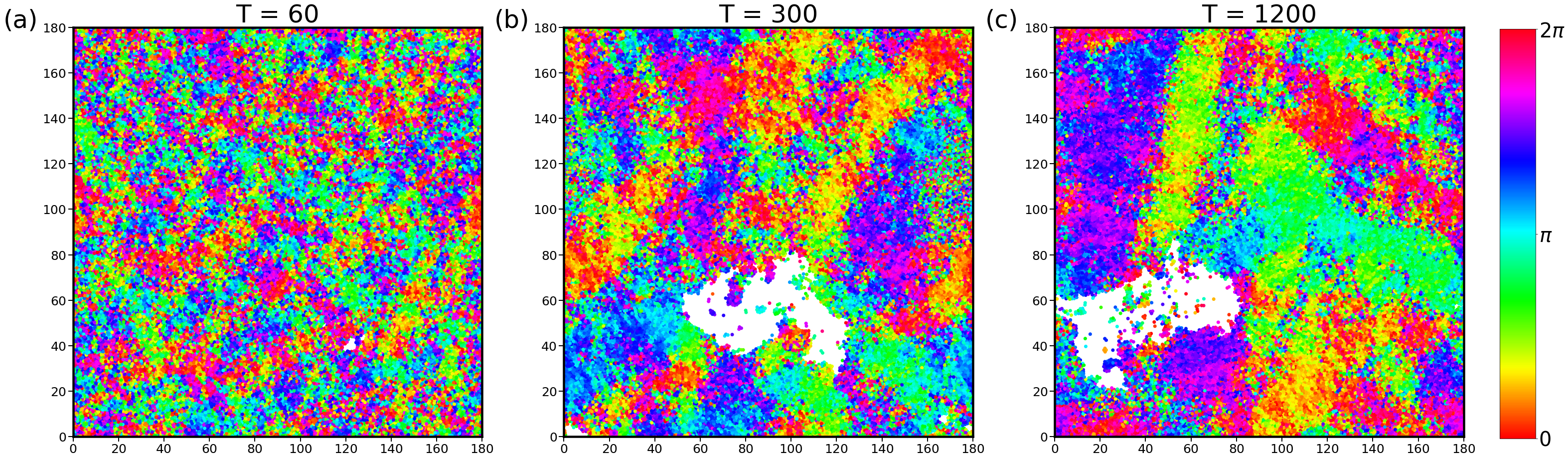}
\includegraphics[width=2\columnwidth]{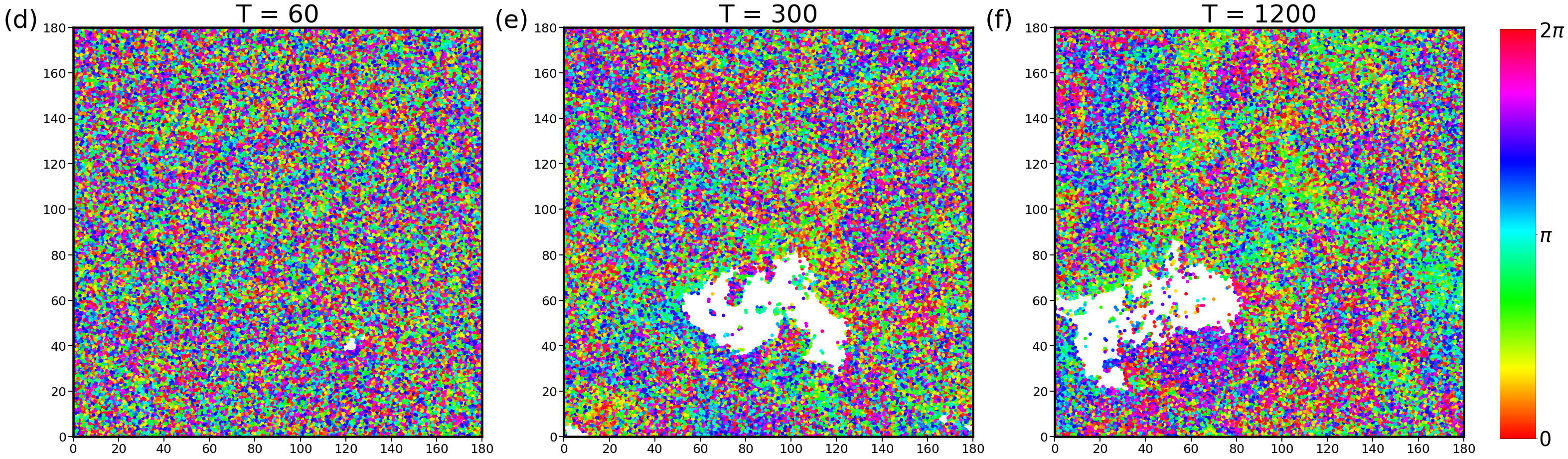}
\caption{Velocity (a)-(c), active force (d)-(f); both active force and velocity heatmaps correspond to the same state of a system of 38880 particles and $L=180$. The color of a dot representing a particle corresponds to the direction of the velocity/active force with the x-axis. The mapping of colors to angles is shown in the color-bar.}
\label{fig:scatterplots}%
\end{figure*}

Fig.~\ref{fig:scatterplots} shows the evolution of a system over time $t$ measured from the instant when the active forces are turned on. In Figs.~\ref{fig:scatterplots}(a)-(c) the colors of the particles correspond to the directions of the velocities. We can see the growth of the velocity domains. In the steady state ($t=1200$) long-range velocity correlations are visible. 
Similarly, the colors of particles in Figs.~\ref{fig:scatterplots}(d)-(f) correspond to the directions of the active forces. The growth of active force domains is visible. Similar results have been reported in Ref.~\cite{suman2025}.

 Apart from the velocity and force correlations apparent from these figures, we also observe the similarity in directions of the velocity and the active force~\cite{suman2025} by comparing figures (a)-(c) to their corresponding figures (d)-(f). This is particularly visible in (c) and (f). 

 In Fig.~\ref{fig:scatterplots}, the snapshots at long times show regions in which very few particles are present. The formation of domains of very low density is a manifestation of motility-induced phase separation~\cite{cates2015} that has been observed in many active systems including the present one~\cite{suman2025}. In our system, the number of particles in the low-density regions is much smaller than that in the high-density region. Therefore, the correlations described below can be interpreted as representing the behaviour of the particles in the high-density region. The occurrence of long-range correlations of the velocity is not contingent upon the formation of voids. Theories of such correlations \cite{henkes2020, szamel2021, caprini2020a, caprini2020b} do not invoke the presence of voids, and Ref.\cite{kuroda2023} shows that such correlations are present even when the formation of voids is suppressed. Our simulations of smaller systems and Ref.\cite{suman2025} show that system-wide velocity correlations and active force correlations are present when the occurrence of voids is not seen clearly. It is, however, possible that the range of these correlations would be reduced~\cite{suman2025} when large voids are present.

 \subsection{Correlation functions: Quantifying the sizes of velocity and force domains}
 To quantify the growth of these domains and understand their morphology, we study the growth of two-point equat-time spatial correlation functions in our system. We compute the self-correlation functions $C_{vv}$, $C_{ff}$ and the cross-correlation function $C_{fv}$. These correlation functions are normalized by their value at $r=1$.  The correlation functions for a configuration obtained at time $t$  are defined as: 
\begin{equation}
    C_{\mathbf{p} \mathbf{q}}(r,t) = \frac{\sum_{i=1}^{N}\sum_{j=i}^{N}(\mathbf{p}_i(t)\cdot\mathbf{q}_j(t))\delta(|{\bf{r}}_i(t)-{\bf{r}}_j(t)|-r)}{\sum_{i=1}^{N}\sum_{j=i}^{N}\delta(|{\bf{r}}_i(t)-{\bf{r}}_j(t)|-r)}
\end{equation}
where $\mathbf{p}$ and $\mathbf{q}$ represent the velocity $\bf{v}$ and the active force $\bf{f}_{a}$. These are averaged over configurations at time $t$ obtained in different runs.

\begin{figure*}
\includegraphics[width=0.65\columnwidth]{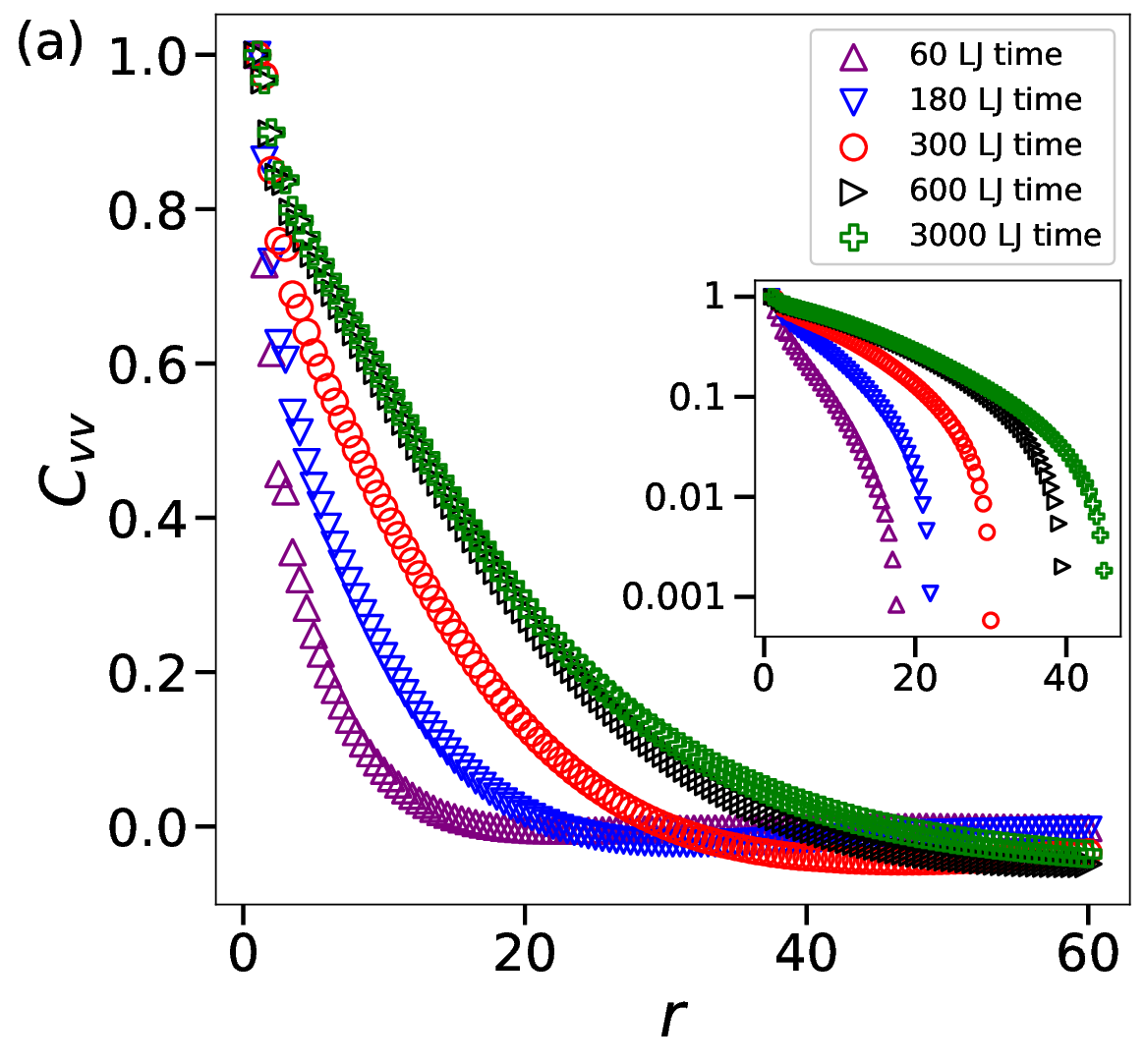}
\includegraphics[width=0.65\columnwidth]{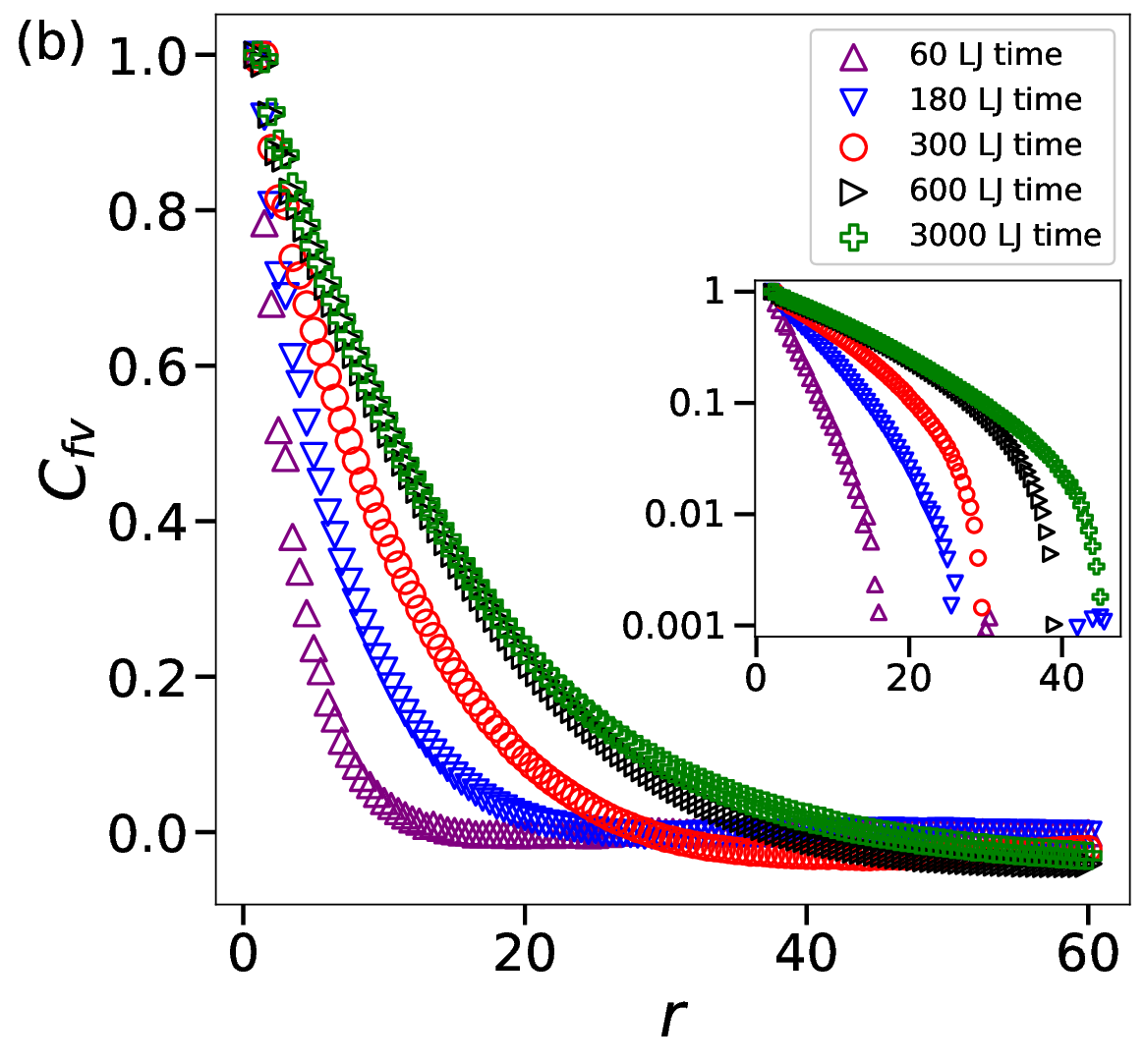}
\includegraphics[width=0.65\columnwidth]{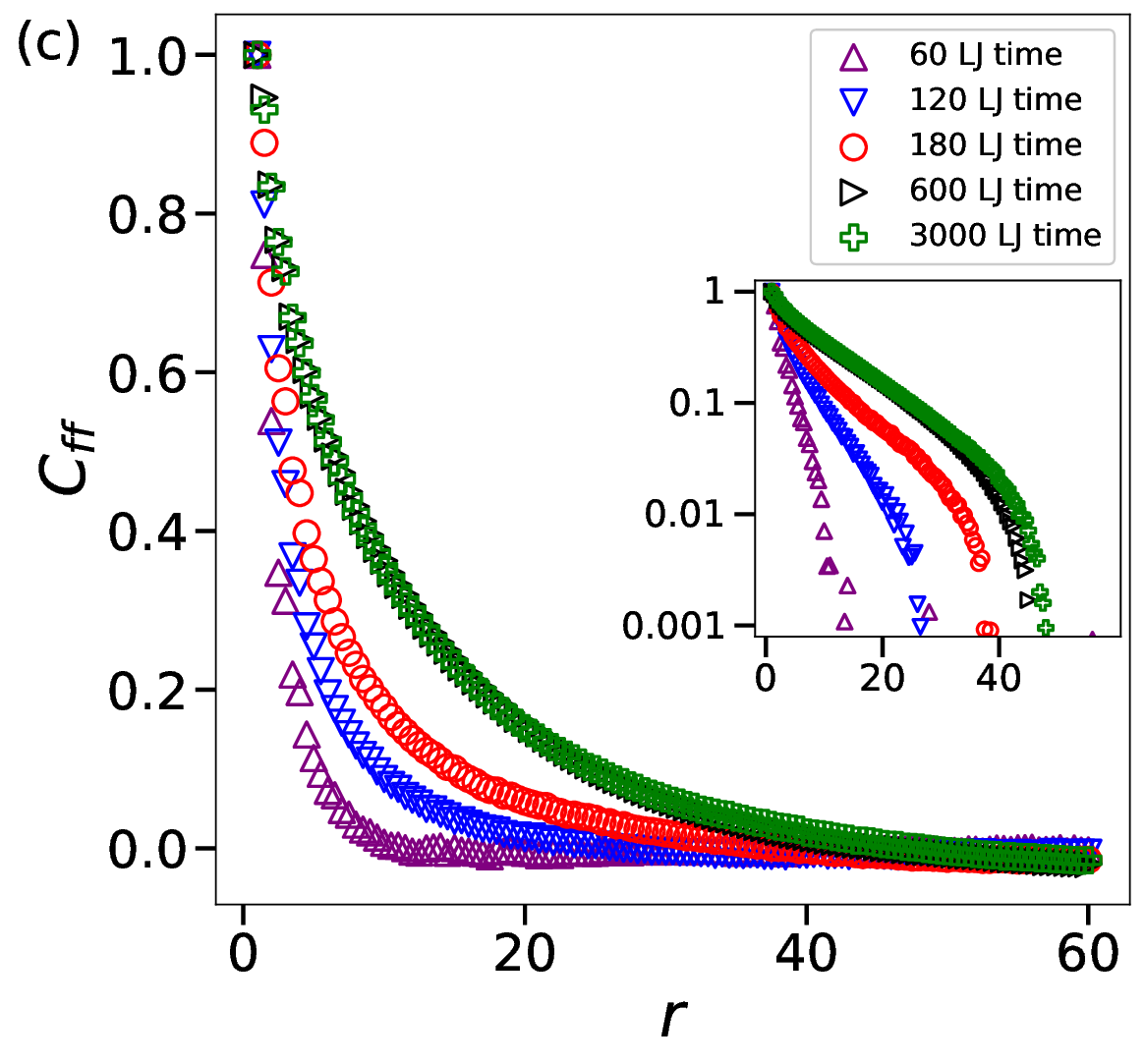}
\caption{The spatial correlation functions in linear scale in the main figure and the log-linear scale in the inset. (a)$~C_{vv}(r,t)$ versus $r$ (b)~$~C_{fv}(r,t)$ versus $r$ (c)~$~C_{ff}(r,t)$ versus $r$}.
\label{fig:spatialcorrelationfunctions}%
\end{figure*}

We plot $C_{vv}, C_{fv}, C_{ff}$ and study their evolution with time $t$ in
Fig.~\ref{fig:spatialcorrelationfunctions}. All three functions grow over time. This is similar to what was seen in Fig.~\ref{fig:scatterplots}. As stated earlier, the cross-correlations between velocity and active force also increase as the system approaches the steady state. The results obtained at large $t$ are consistent with those reported in Ref.~\cite{suman2025}.

To quantify the size of these domains we use $R(t)$, defined as the distance over which the correlation function decreases to 0.1 of its maximum value i.e.~$C(R(t),t) = 0.1$.
If the growth of the domains is self-similar in the sense that the morphology of the domains remains the same, and there is only an increase in the characteristic size of such domains with increasing time, the correlation functions should exhibit a dynamical scaling relation with the characteristic domain size~$R(t)$, i.e. the correlation functions should be independent of time when $r$ is scaled by~$R(t)$
\begin{equation}
\label{eq:scaling_corr}
C(r,t) = g\left(\frac{r}{R(t)}\right).
\end{equation}

\begin{figure}
\includegraphics[width=0.45\columnwidth]{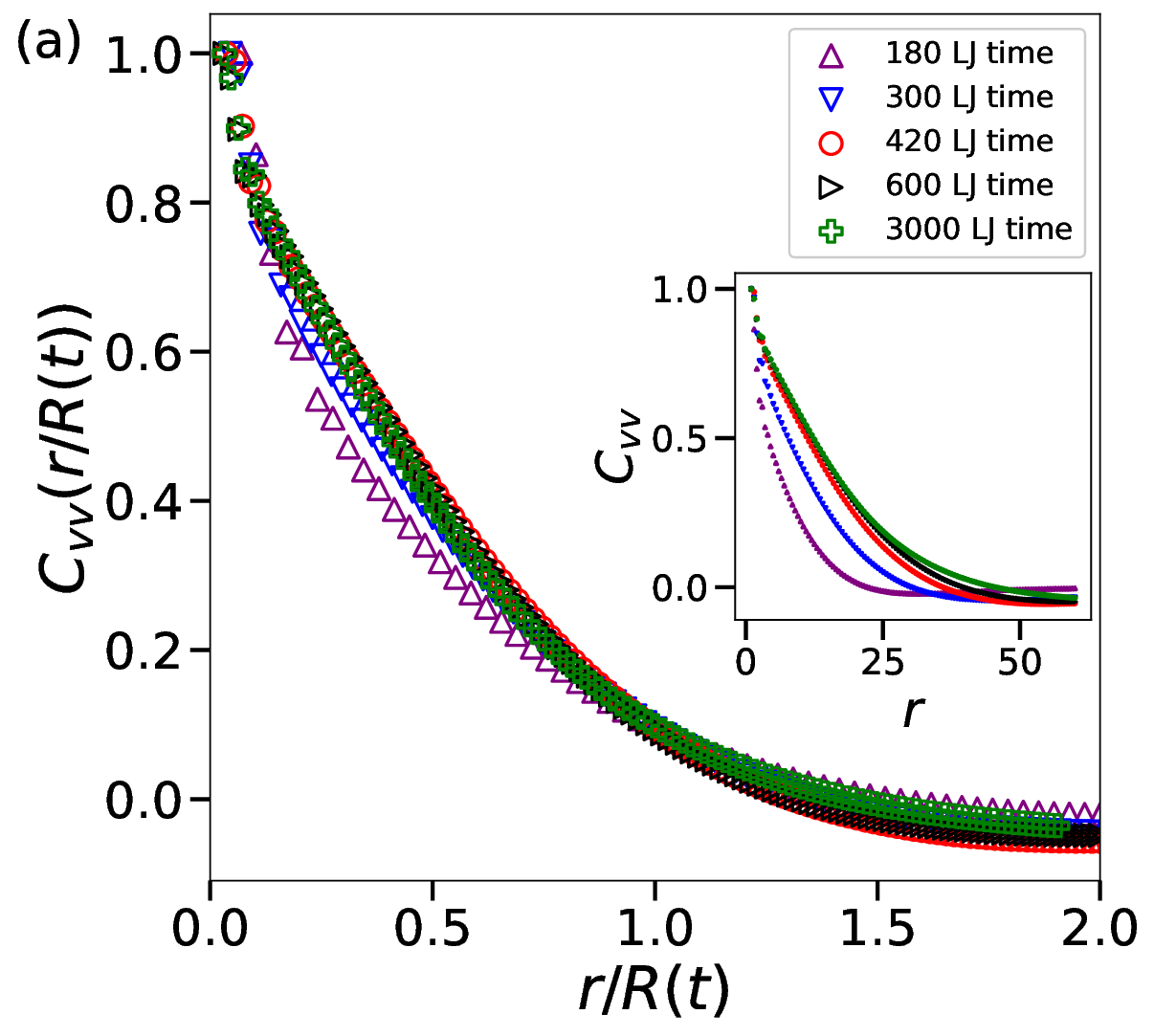}
\includegraphics[width=0.45\columnwidth]{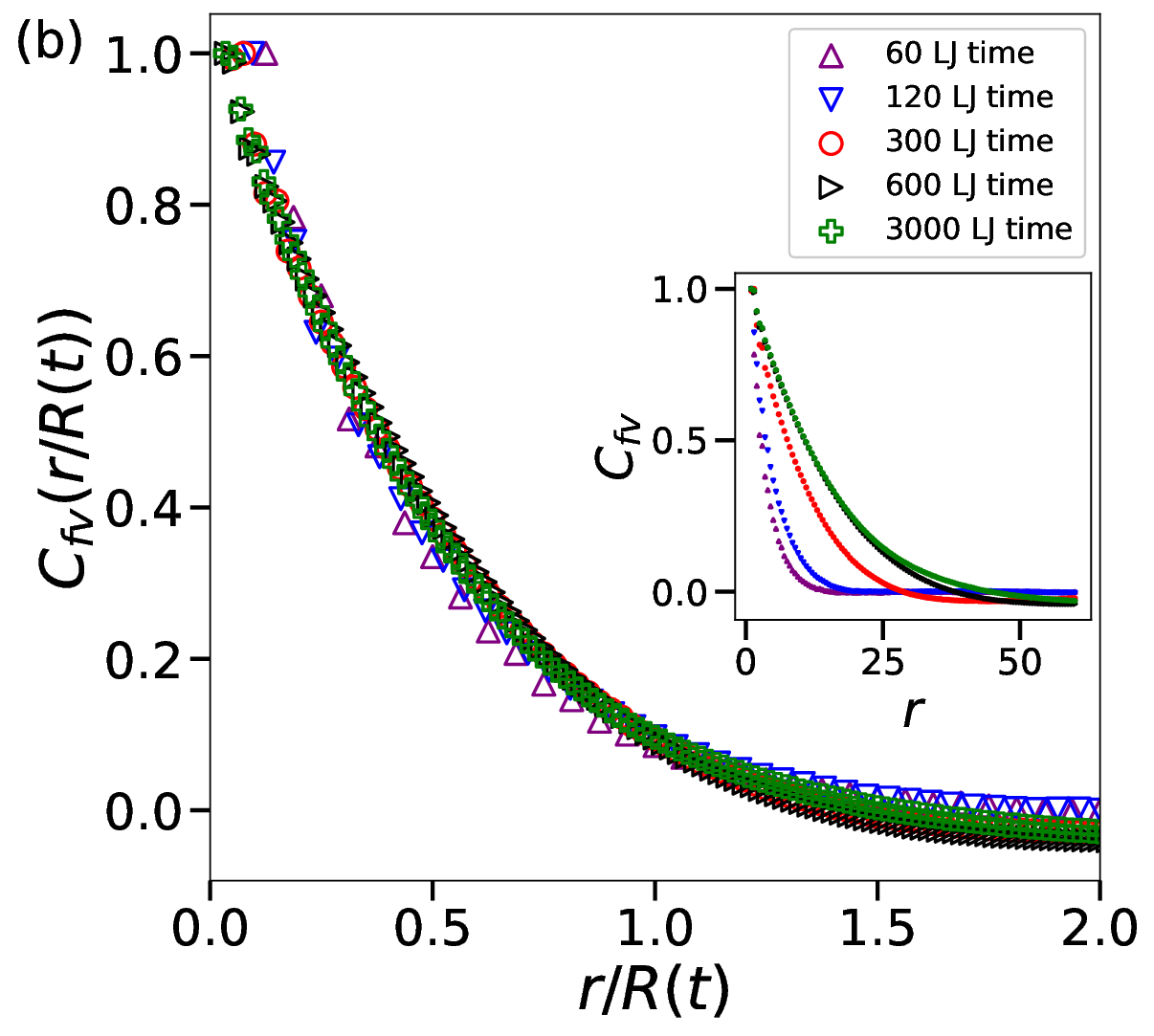}
\includegraphics[width=0.45\columnwidth]{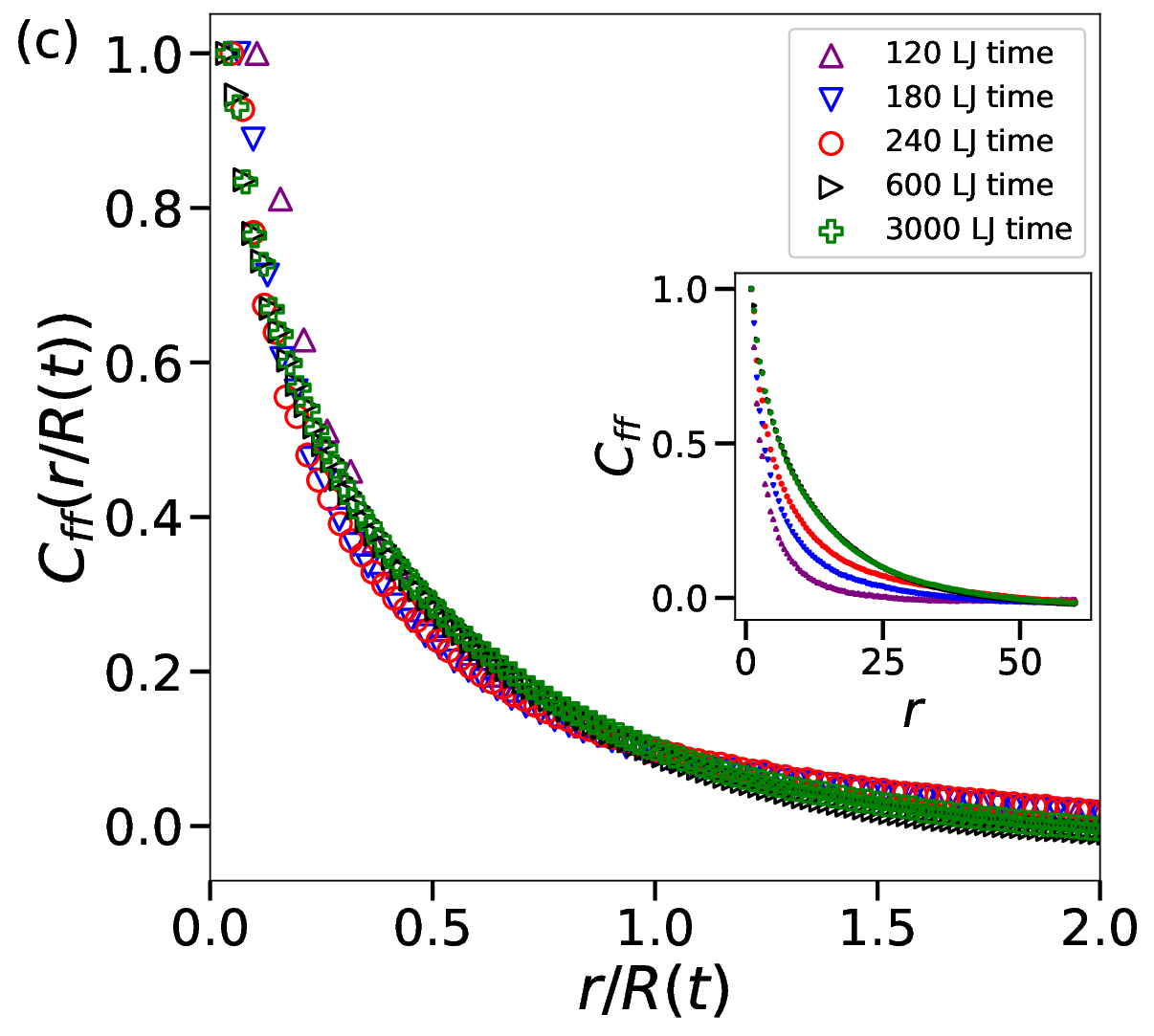}
\includegraphics[width=0.45\columnwidth]{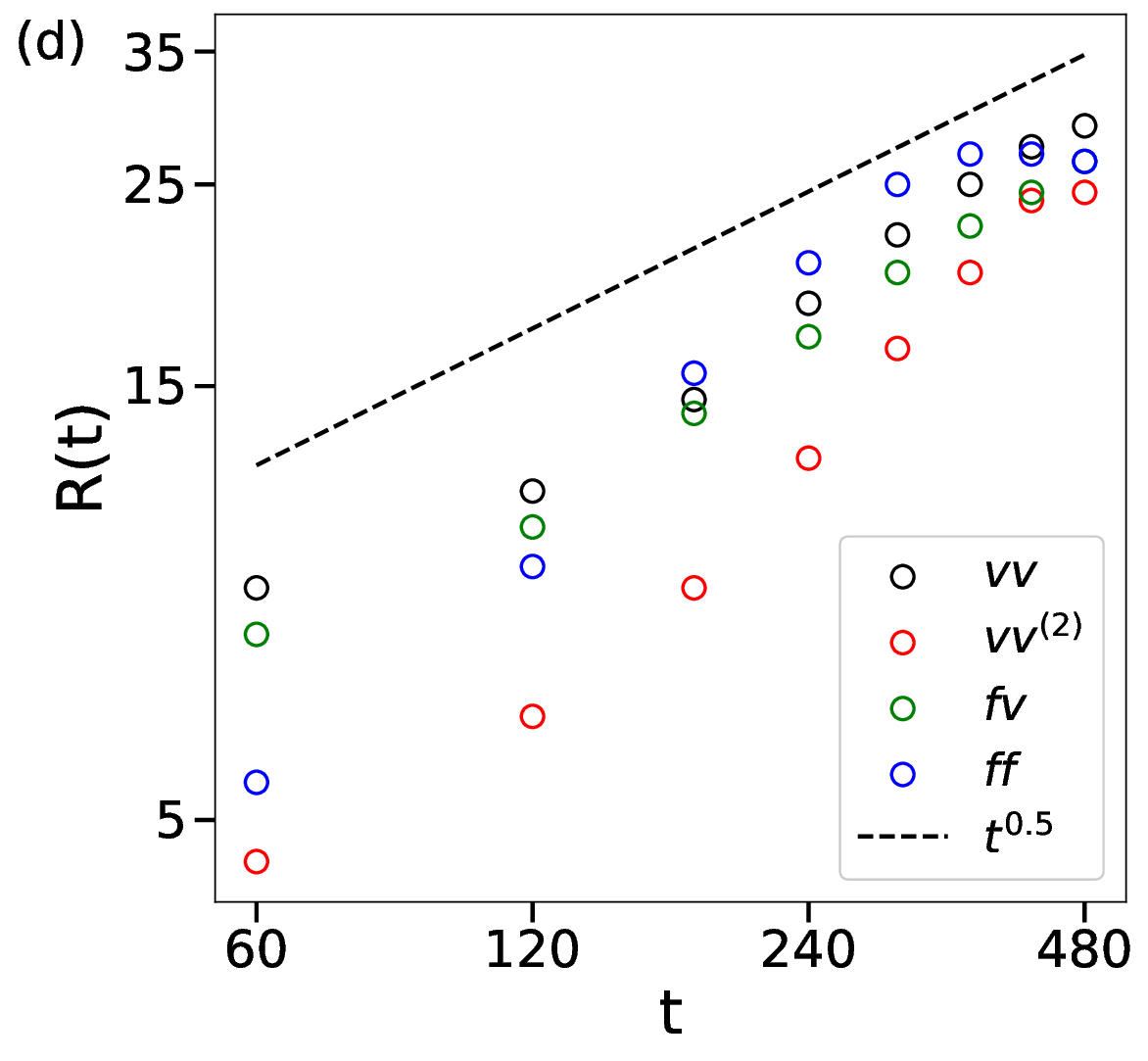}
\caption{Spatial correlation functions plotted versus the scaled distance $r/R(t)$ at different times. Unscaled versions are shown in the insets (a)~$C_{vv}\left(\frac{r}{R(t)}\right)$ versus $\frac{r}{R(t)}$ (b)~$C_{fv}\left(\frac{r}{R(t)}\right)$ versus $\frac{r}{R(t)}$ (c)~$C_{ff}\left(\frac{r}{R(t)}\right)$ versus $\frac{r}{R(t)}$ (d)~$R(t)$ as a function of time $t$ for $C_{vv}$, $C_{fv}$, $C_{ff}$ and $C_{vv}^{(2)}$ on a log-log scale.}
\label{fig:rescaledspatialcorrelationfunctions}%
\end{figure}

We present the plots for the correlation functions versus the scaled distance and~$R(t)$ as a function of $t$ for the three types of spatial correlation functions under consideration in Fig.~{\ref{fig:rescaledspatialcorrelationfunctions}. 
 In all three cases, the correlation functions at different times exhibit reasonable collapses to a single function $g(r/R(t))$, except possibly the data for the earliest time for $C_{vv}$. This confirms that the growth of correlations in this system is self-similar. 

 The time-dependence of the length scale $R(t)$ for the three correlation functions is shown in Fig.\ref{fig:rescaledspatialcorrelationfunctions}(d). It is clear that the three length scales grow with time in a similar way.
 In usual domain growth processes, typically $R(t)\approx t^{\delta}$ where $\delta$ is the growth exponent~\cite{bray2002}. Since the range of growth of the correlation lengths is about one decade in our simulations, it is not possible to obtain accurate estimates of the growth exponents. However, we can see in
Fig.\ref{fig:rescaledspatialcorrelationfunctions}(d) that for all three correlation functions, $\delta \approx 0.5$. As noted in section~\ref{sec:introduction}, the symmetry of the ordering fields is that of the XY model and the ordering fields are conserved. The domain growth exponent for the conserved XY model in two dimensions is expected to be $\delta = 0.25$ \cite{bray1989,bray1990,mondello1993}. Thus, we can conclude that the domain growth exponent in our strongly non-equilibrium system is different from that in equilibrium models with the same symmetry.

It is apparent from the plots in the insets of Fig.\ref{fig:spatialcorrelationfunctions} that the initial part of the decay of the correlation functions with increasing $r$ can be approximated by an exponential function, $C(r,t) \sim \exp(-r/\xi(t))$. The time-dependence of the length scale $\xi(t)$ and another length scale $R_0(t)$ defined as the value of $r$ at which $C(r,t)$ crosses zero is similar to that of $R(t)$. This is shown in Fig.\ref{fig:lengths} in Appendix II.

\subsection{Structure factors}
Now we know that the domains grow in time in a self-similar way. If we can ascertain the nature of the boundaries of these domains, we will have complete information about them.
For this we make use of the structure factor:
\begin{equation}
    S_{pp}({\bf{k}},t) = \langle{\bf{p}}({{\bf{k}},t})\cdot{\bf{p}}(-{\bf{k}},t)\rangle
\end{equation}
Here, $\bf p$ represents the velocity $\bf v$ or the active force ${\bf{f}}_a$, $\langle\cdots\rangle$ represents an average over different runs and
\begin{equation}
    {\bf{p}}({\bf{k}}, t) = \frac{1}{N}\sum_{j=1}^{N} \exp[i {\bf{k}} \cdot {\bf{r}_j}(t) ] {\bf{p}}_j(t)  
\end{equation}
We circularly average $S_{pp}({\bf{k}},t)$ over $\bf{k}$ to obtain  $S_{pp}(k,t)$ where $k = |{\bf{k}}|$. In systems with sharp boundaries between domains, the $k$-dependence of the structure factors for large $k$ ($kL \gg 1)$ is given by Porod's law~\cite{porod83}: $S(k) \propto 1/k^{(d+1)}$ where $d$ is the spatial dimension. This corresponds to the correlation function in real space behaving as $C(r) \sim C(0) - Ar$ near $r=0$ where $A$ is a positive constant. Since $d=2$ in our system, $S(k) \propto 1/k^3$ corresponds to agreement with Porod's law. In systems with rough domain boundaries, $C(r) \sim C(0) - Ar^\alpha$ and $S(k) \propto 1/k^{d+\alpha}$ with $\alpha<1$.

We plot the structure factors for our system in Fig.~({\ref{fig:structurefactors}). For velocity domains, the structure factor in the steady state has good agreement with $S_{vv}\approx k^{-3}$. This signifies the existence of smooth boundaries of the velocity domains. This is in agreement with the result found in Ref.~\cite{suman2025}.
  For the active-force domains, the structure factor decays with an exponent close to 2, suggesting that $\alpha \sim 0$. This signifies the existence of rough boundaries between the force domains. The domain morphologies shown in Fig.~\ref{fig:scatterplots} are qualitatively consistent with this conclusion.
\begin{figure}
\includegraphics[width=0.45\columnwidth]{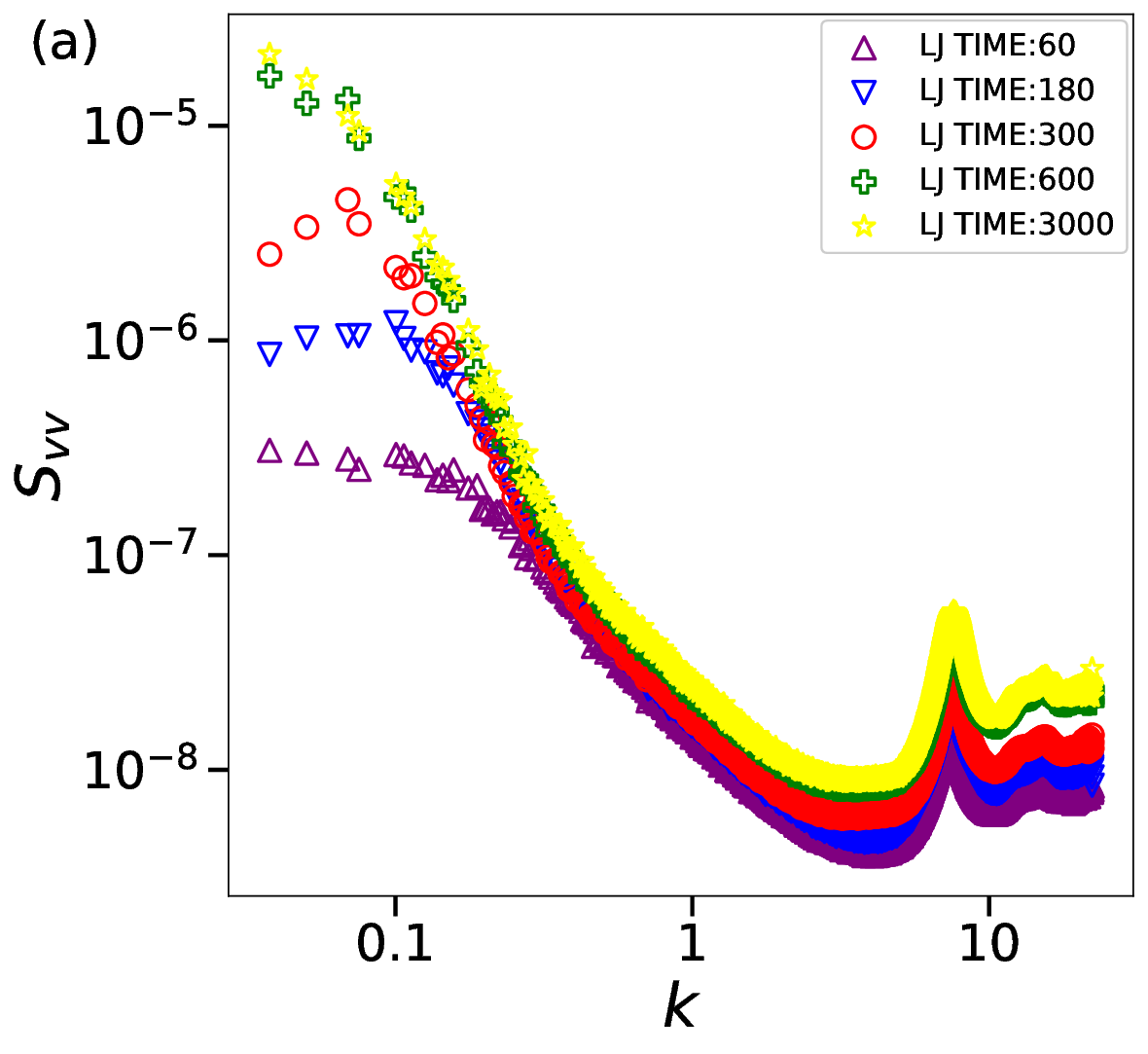}
\includegraphics[width=0.45\columnwidth]{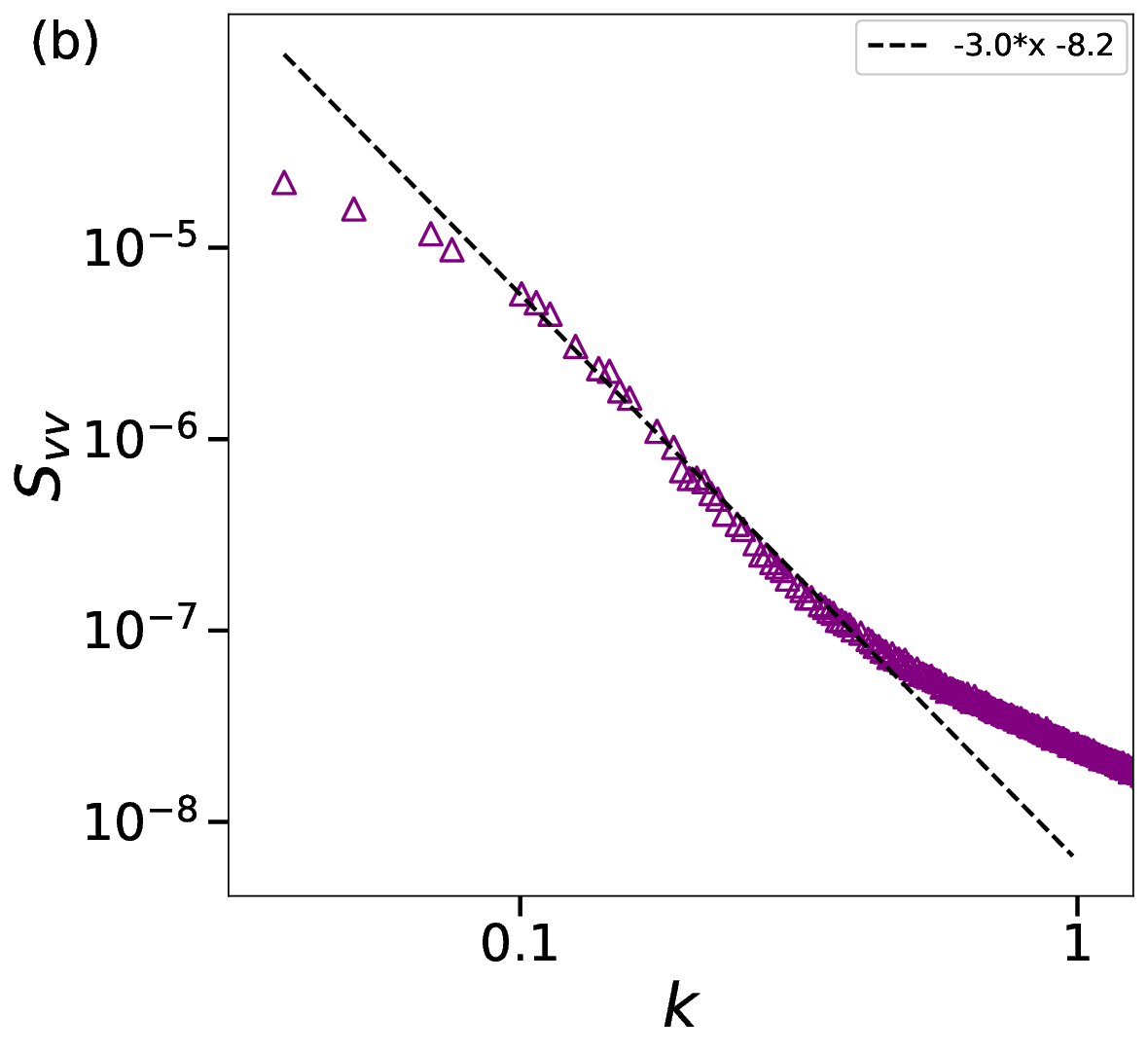}
\includegraphics[width=0.45\columnwidth]{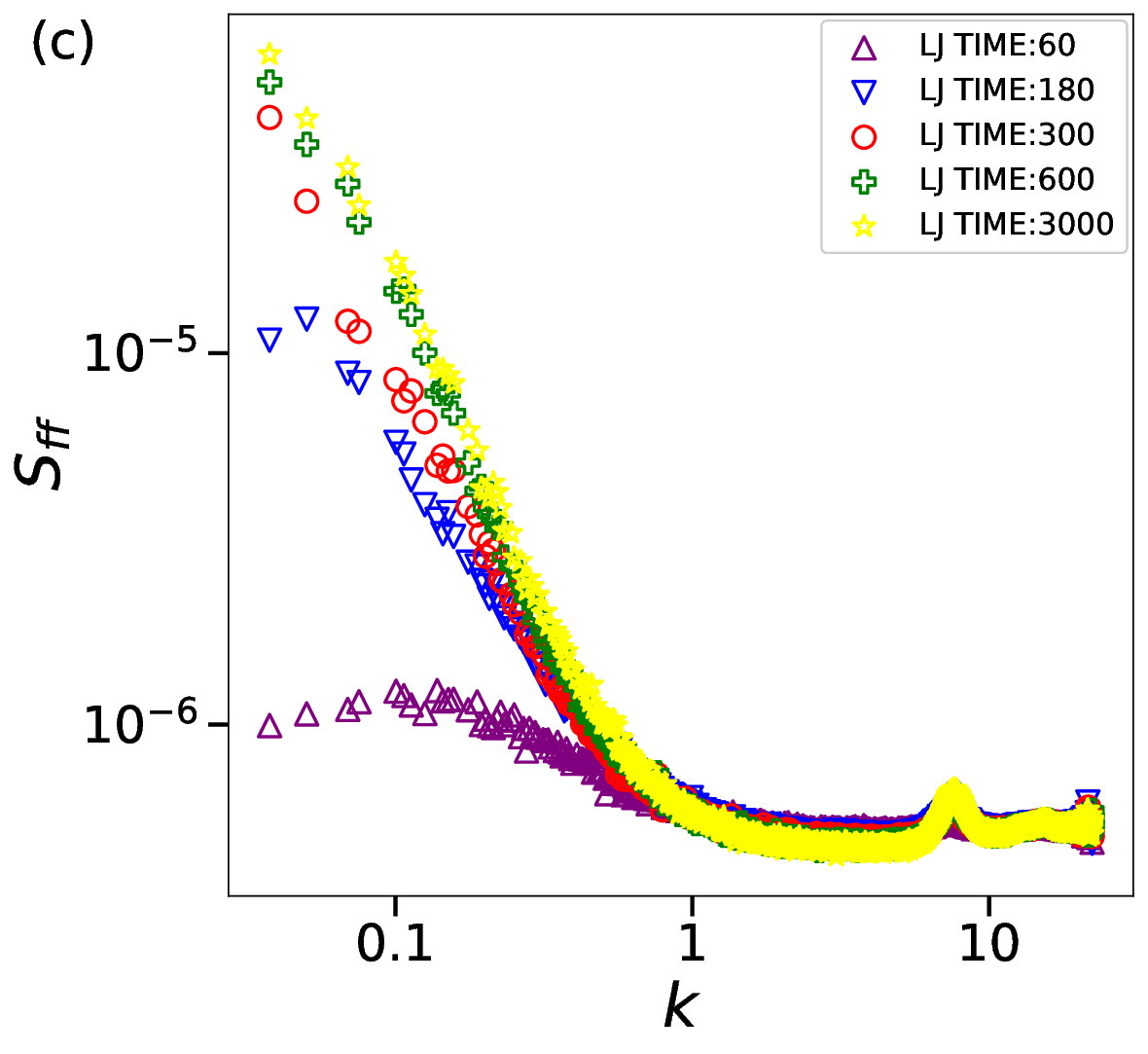}
\includegraphics[width=0.45\columnwidth]{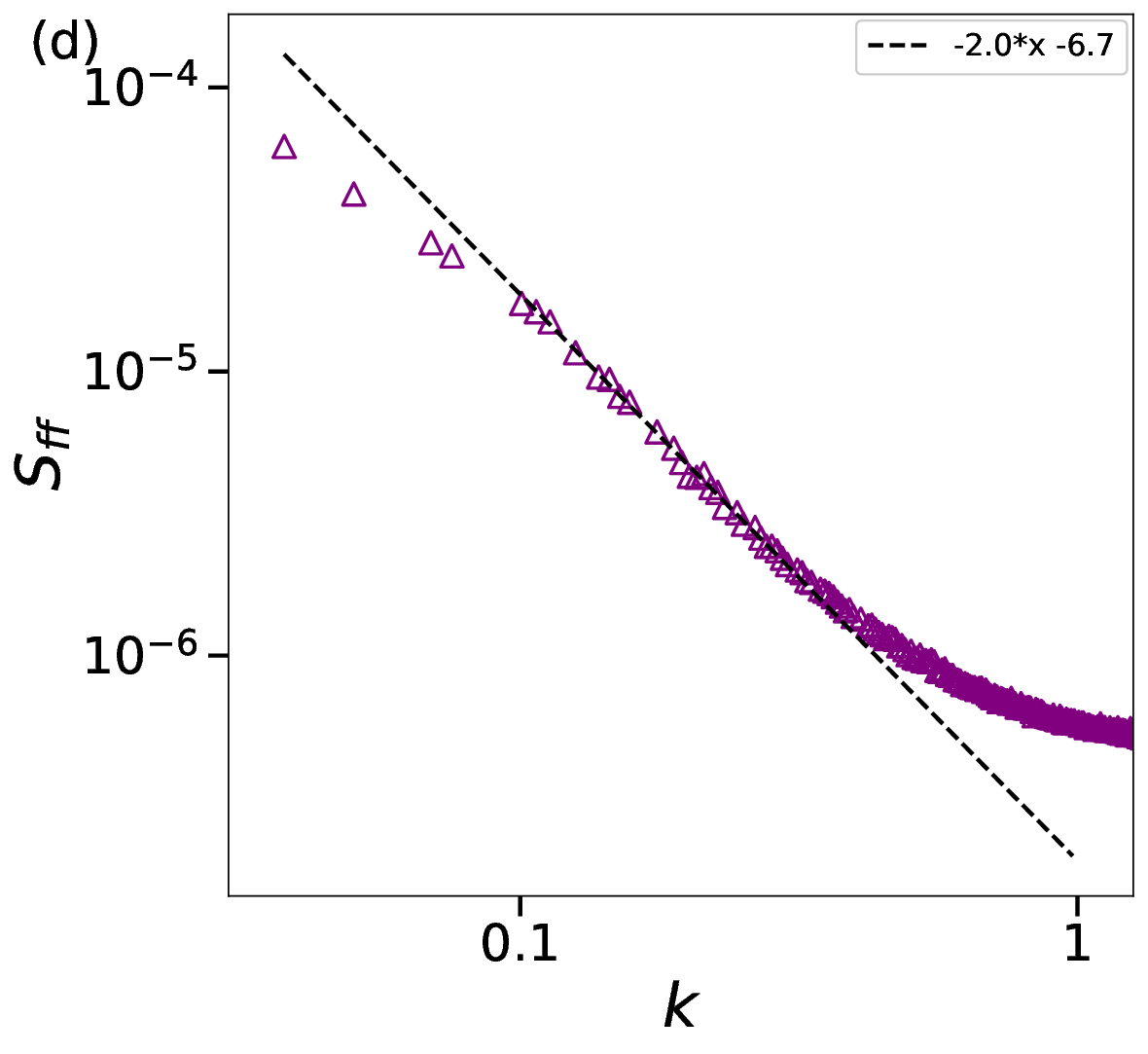}
\caption{The structure factors $S(k,t)$ plotted versus $k$ on log-log scale. (a),(b) $~S_{vv}(k,t)$ versus $k$ in the steady state and (c),(d)~$S_{ff}(k,t)$ versus $k$ in the steady state. The dashed lines in (b) and (d) represent power laws of the form $y=cx^m$ where the exponents are given in the legend.}
\label{fig:structurefactors}%
\end{figure}

\subsection{Anisotropy in the flow pattern}

In our system, the interparticle interactions are isotropic and the directions of the active forces are distributed uniformly between $0$ and $360$ degrees. Therefore, the directions of the velocities of the particles in the steady state are expected to be uniformly distributed between $0$ and $360$ degrees.
Surprisingly,
on plotting the number of particles \textit{versus} the angle $\theta$ of the velocity with the x-axis for a randomly chosen steady-state snapshot of our system in Fig.\ref{fig:lambdaphi}(a), we notice well-defined peaks in two directions that are opposite to each other (separated by 180 degrees). The coarse-grained velocity field in this configuration is shown in Fig.\ref{fig:phi_quiver} in which coarse-grained velocities with magnitude smaller than 0.3 are not shown for clarity. In this figure, the length and the direction of an arrow represent, respectively, the magnitude and the direction of the coarse-grained velocity at the location of the arrow. The presence of two large groups of particles moving in the $+x$ ($\theta = 0$ ) and $-x$ ($\theta$ = 180 degrees) are clearly seen in this figure.

\begin{figure}
\includegraphics[width=0.99\columnwidth]{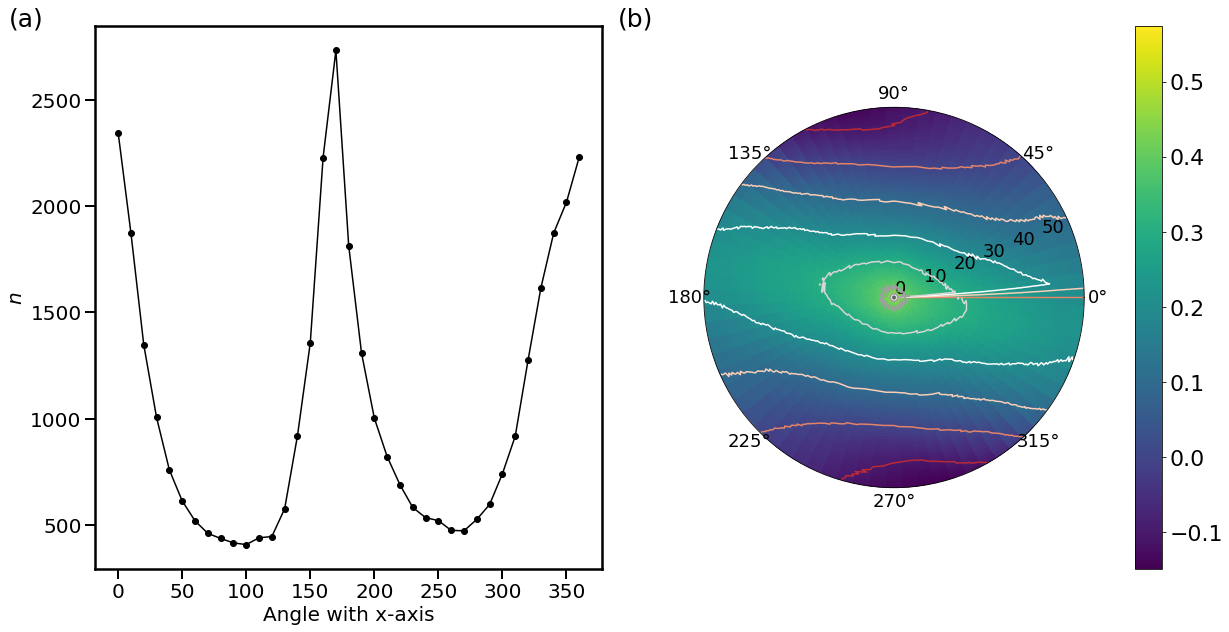}
\caption{(a)~Distribution of velocity angles, (b)~Contour plot for $C_{vv}(r,\Theta)$ 
of the same configuration in the steady state.}
\label{fig:lambdaphi}%
\end{figure}

\begin{figure}
\includegraphics[width=0.79\columnwidth]{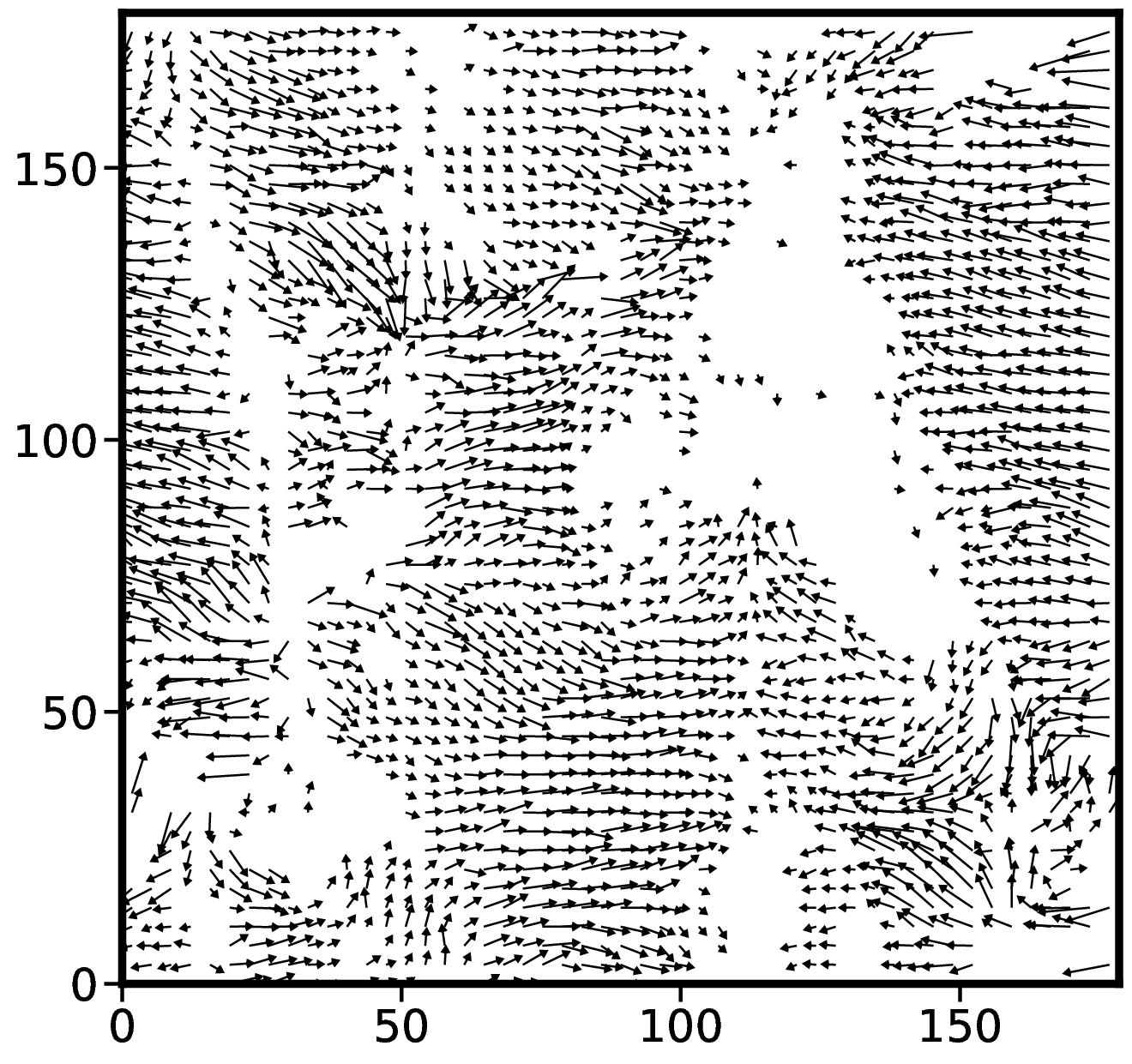}
\caption{Velocity field for the coarse-graining scale $\Omega$ = 3.5 and $|v|>0.3$ for the configuration in Fig.\ref{fig:lambdaphi}.}
\label{fig:phi_quiver}%
\end{figure}

To explore this further, we define a modified correlation function as:
\begin{equation}
    C_{\mathbf{p} \mathbf{q}}(r,t,\Theta) = \frac{\sum_{i}\sum_{j}(\mathbf{p}_i(t)\cdot\mathbf{q}_j(t))\delta(|{\bf{r}}_{ji}(t)|-r)\delta(\Theta_{r_{ji}}-\Theta)}{\sum_{i}\sum_{j}\delta(|{\bf{r}}_{ji}(t)|-r)\delta(\Theta_{r_{ji}}-\Theta)}
\end{equation}
Here, $\Theta_{r_{ji}}$ is defined as the angle made by the relative vector ${\bf{r}_{ji}}$ = ${\bf{r}}_j-{\bf{r}}_i$ with the x-axis. $\mathbf{p}$ and $\mathbf{q}$ have the same meaning as defined earlier.
We plot this correlation function $C_{vv}(r,\Theta)$ for the same configuration in Fig.\ref{fig:lambdaphi}(b) in which the color at a point $(r,\Theta)$ represents the value of $C_{vv}(r,\Theta)$ according to the color-bar shown next to the plot. Contours of constant $C_{vv}(r,\Theta)$ are also shown.  The tendency of the particle velocities to point along specific direction should show up as a departure from isotropy in $C_{vv}(r,\Theta)$. If the motion of the particles is dominated by two streams moving in directions $\theta$ and $\theta+\pi$, then two particles $i$ and $j$ for which $\Theta_{r_{ji}}$ is close to $\theta$ or $\theta+\pi$ are likely to have their velocities in the same direction and $C_{vv}(r,\Theta)$ would be larger than its values for other directions of $\Theta$. As shown in Fig.\ref{fig:lambdaphi}(b), such anisotropy is present in the dependence of $C_{vv}(r,\Theta)$ on the angle $\Theta$. We classify this tendency as anisotropy in the velocity distribution. This anisotropy is approximately uniaxial in the sense that the distribution has peaks in two directions that are opposite to each other. We also find a tendency of these directions to align with the axes corresponding to the sides of the simulation box ($\theta$ = 0 (180) degrees and 90 (270) degrees). 

To quantify this anisotropy we consider a quantity that is used to define the degree of alignment (nematic order) in rod-like objects with uniaxial symmetry. In this analogy, the direction of the velocity of a particle plays the role of the direction of the long axis of a rod-like molecule. Let $\theta_i$ be the angle that the velocity ${\bf v}_i$ of particle $i$ makes with the x-axis. We define
\begin{equation}
    A = \sum_i^N \frac{\cos^2\theta_i}{N} - \frac{1}{2}
\end{equation}
\begin{equation}
    B = \sum_i^N \frac{\cos\theta_i \sin\theta_i}{N}.
\end{equation}
Here, $N$ is the number of particles. We find the eigenvalues and eigenvectors of the following matrix 
\[
\begin{bmatrix}
    $A$       &$B$ \\
    $B$       &$-A$  \\
\end{bmatrix}.
\]
The positive eigenvalue is $\lambda = \sqrt{A^2 + B^2}$ and the corresponding normalized eigenvector is 
\begin{equation}
    \alpha = \frac{B}{2\lambda(\lambda-A)},
    \beta = \frac{\sqrt{\lambda-A}}{\sqrt(2\lambda)}.
\end{equation}
 The eigenvector elements $\alpha$ and $\beta$ define an angle $\phi~=~\tan^{-1}(\beta/\alpha)$. This procedure defines a number $\lambda$ and an angle $\phi$ for each configuration. $\lambda$
provides a measure of the strength of the anisotropy in the distribution
 of the angles ${\theta_i}$ ($\lambda$ = 0 if the distribution is uniform between 0 and 2$\pi$ and $\lambda = 0.5$ for perfect alignment), and the distribution is expected to show peaks at the angles $\phi$ and $\phi$+$\pi$.

\begin{figure}
\includegraphics[width=0.99\columnwidth]{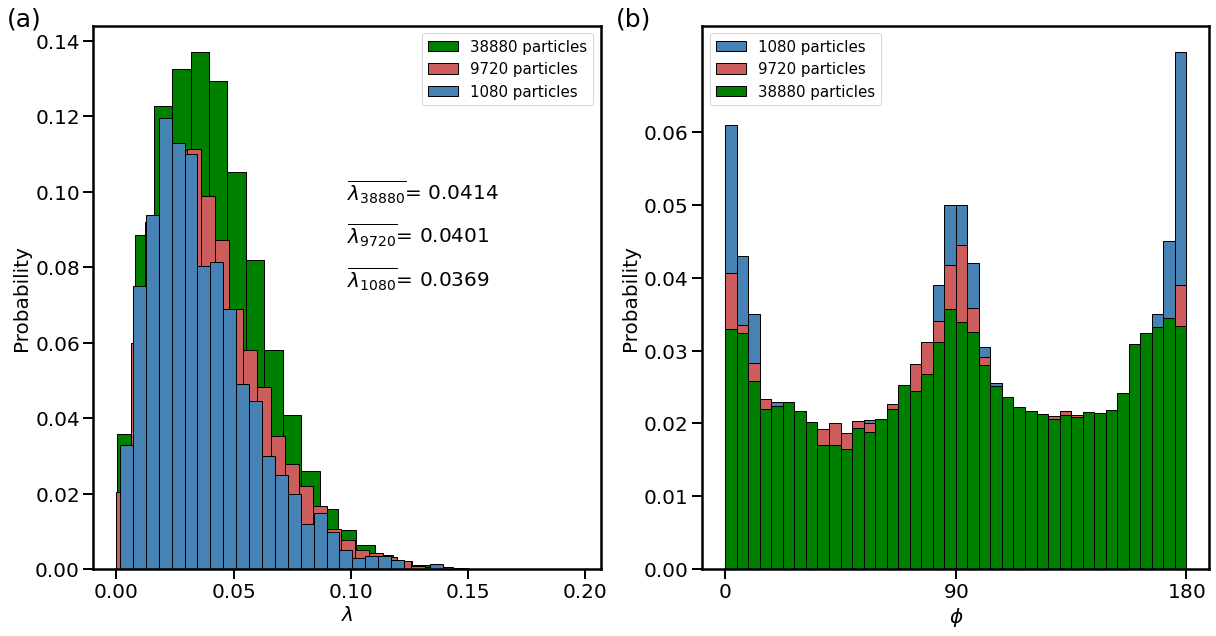}
\caption{(a)~Histogram of $\lambda$ and (b)~Histogram of $\phi$}
\label{fig:histograms}%
\end{figure}
 We plot the distributions of $\lambda$ and $\phi$ for three system sizes in the steady state in Fig.~\ref{fig:histograms}. We average over 2000 steady-state configurations for the system of 1080 particles, 17000 configurations for the system of 9720 particles and 36000 configurations for 38880 particles.
 We find that for all system sizes $\lambda \in [0,0.15]$, and the mean value increases as we move to larger systems. This implies that the anisotropy of the velocity distribution is not a finite-size effect. The distribution of $\phi$ in Fig.~\ref{fig:histograms}(b) shows distinct peaks along the directions of the boundaries of our simulation box i.e. the x and y axes. However, the heights of the peaks decrease with increasing $N$, suggesting that the tendency of the preferred direction of the velocities to align with the boundaries of the simulation box is a finite-size effect that would disappear in the large-$N$ limit. To check this further, We also calculate the value of $\langle\cos(4\phi)\rangle$ for all three system sizes. The values are 0.00172, 0.00128 and 0.000156 for systems of 1080, 9720 and 38880 particles, respectively. As expected from Fig.\ref{fig:histograms}(b) this quantity decreases with increasing $N$ - as the system gets larger, the motion is more likely to be along any direction.  From these observations, we conclude that in the steady state of our system, the motion of the particle is dominated by two streams traveling in opposite directions and the preferred direction changes randomly with time. 

 To examine further the dynamics of the preferred direction $\phi$, we have calculated the autocorrelation function 
 \begin{equation}
     C_{\phi\phi}(t) = \langle \cos2(\phi(t)-\phi(0))\rangle,
 \end{equation}
 in the steady state. Here, $\langle \cdots \rangle$ represents an average over initial time and different runs, and the correlation function is defined to take into account the fact that $\phi$ and $\phi +\pi$ are equivalent. The results for this correlation function are shown in Fig.\ref{fig:phi}(a). It decays to zero at long times, implying that the preferred direction of flow gets randomized in the steady state, and rotational symmetry is not broken.

 In view of the tendency of the particles to move in two oppositely directed streams, it is interesting to consider a modified spatial correlation function of the particle velocities in which the angle $\theta$ that the velocity of a particle makes with the $x$-axis and $\theta + \pi$ are taken to be equivalent. This correlation function is defined as 
 
 \begin{align}
     C_{vv}^{(2)}(r,t) &=\sum_{i=1}^{N}\sum_{j=i}^{N}(|\mathbf{v}_i(t)||\mathbf{v}_j(t)|\cos2(\theta_i(t)-\theta_j(t)))\nonumber \\
     &\times \delta\left(|{\bf{r}}_i(t)-{\bf{r}}_j(t)|-r\right)\nonumber\\
 &/\left[{\sum_{i=1}^{N}\sum_{j=i}^{N}\delta(|{\bf{r}}_i(t)-{\bf{r}}_j(t)|-r)}\right]
 \end{align}
Here, $\theta_i(t)$ is the angle of the velocity ${\bf{v}}_i(t)$ of particle $i$ at time $t$ with the $x$-axis. Results for this correlation function at different times are shown in Fig.\ref{fig:phi}(b). As expected, the decay of the correlations with distance becomes slower as time progresses. The main difference between the results shown here and those in Fig.\ref{fig:spatialcorrelationfunctions}(a) is that the negative values of $C_{vv}(r,t)$ for large $r$ are not found for $C_{vv}^{(2)}(r,t)$. This is because values of $(\theta_i - \theta_j)$ near $\pi$ lead to negative values of $C_{vv}$ for large $r$, but this does not happen for $C_{vv}^{(2)}$. As shown in the inset of Fig.\ref{fig:phi}(b), the growth of these correlations with time exhibits dynamic scaling. The corresponding length scale $R$ is shown as a function of time in Fig.\ref{fig:rescaledspatialcorrelationfunctions}(d). It is clear from the plots that this length scale grows faster in time than those obtained from the other correlation functions. Fitting the growth of this length to a power law, we get a growth exponent $\delta$ close to 0.8, substantially larger than the exponent $\delta = 0.5$ expected~\cite{bray2002} for non-conserved spin models.

\begin{figure}
\includegraphics[width=0.96\columnwidth]{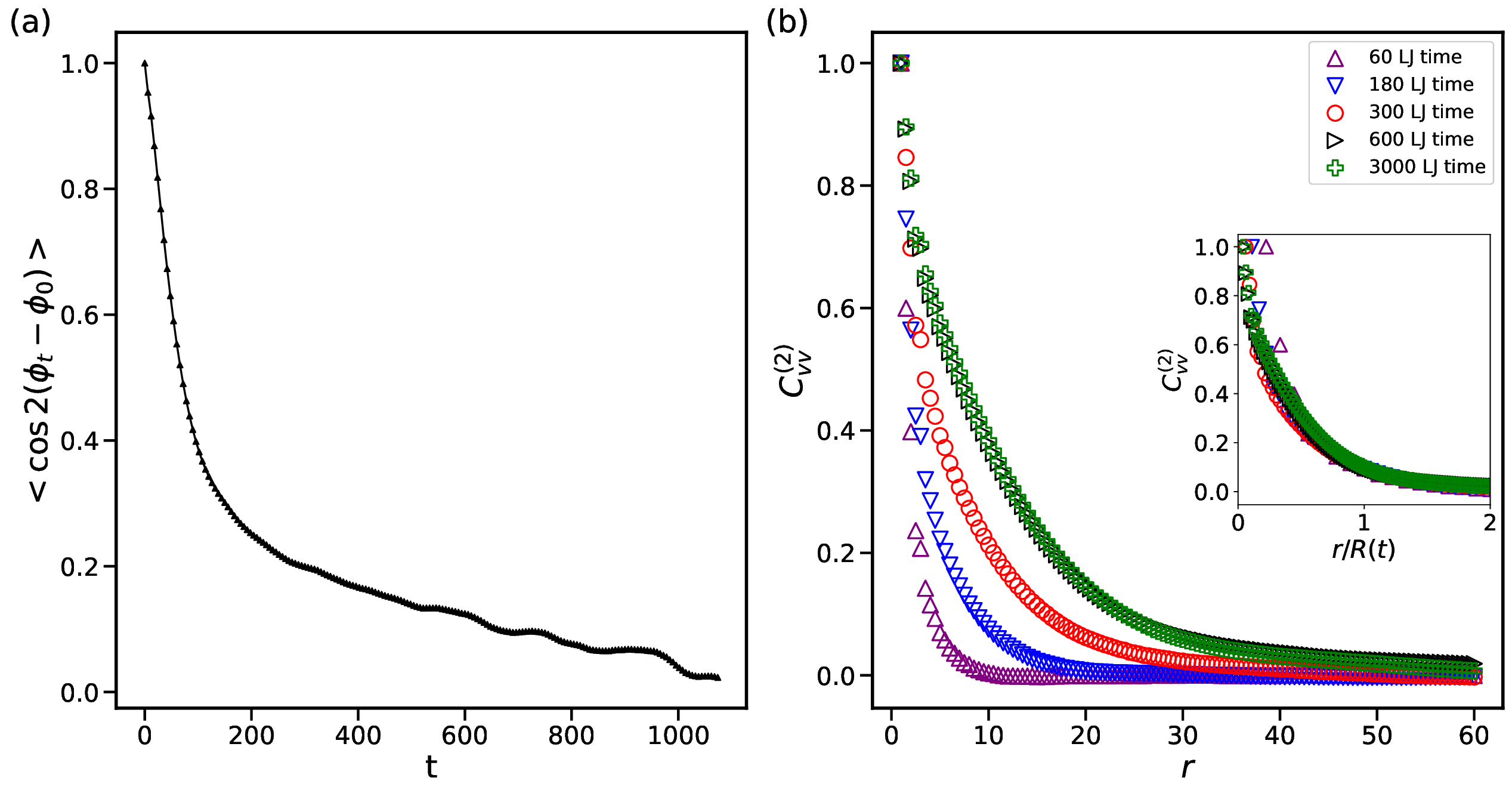}
\caption{(a)~Auto-correlation of $\phi$ and (b) The modified spatial correlation function $C_{vv}^{(2)}$ versus $r$ for different times in the main figure and the scaled correlation function $C_{vv}^{(2)}\left(\frac{r}{R(t)}\right)$ versus $\frac{r}{R(t)}$ in the inset.}
\label{fig:phi}%
\end{figure}

\section{Summary and Discussion}
\label{conclusions}

In this work, we have studied by simulations the process of development of spatial correlations of particle velocities and active forces in the self-organized liquid state of an athermal two-dimensional system of particles with persistent active forces. This process is similar to the growth of order in a system with an order-disorder transition after a quench from a disordered state to a state in the ordered region of the phase diagram. Our system exhibits self-similar growth of correlations with a growth exponent that is substantially different from the growth exponent in equilibrium systems with the same symmetry. The boundaries of velocity domains are found to be smooth, whereas those of domains of active forces are rough. Analysis of the flow pattern in the steady state shows evidence for the flow to be dominated by two streams of particles moving in opposite directions.\\

Experimental realizations of particles with active forces that remain constant in time irrespective of the position or the orientation of the particles is difficult. If the persistence time of the active forces in an experimental system is substantially larger than the time taken by our system to reach a steady state, then some of the phenomena seen in the steady state of our system are expected to be observed in the liquid state of the experimental system. It would be interesting to check this in experiments and simulations. The time to reach the steady state in the largest system considered in our simulations is about 1000 time units. It should be possible to simulate systems of active particles with persistence times that are substantially larger than this value. The time to reach the steady state increases with the system size $N$. Our limited investigation of the system-size dependence suggests that the time to reach the steady state is proportional to $N^\alpha$ with $\alpha \simeq 0.43$. This observation implies that the persistence time must be
scaled appropriately with the system size for the effects seen in the present study to be observed in
simulations of systems with finite persistence time. In any case, the system we have studied is a simple one that exhibits interesting self-organization that deserves detailed investigation on its own merit.

We close with a discussion of possible theoretical explanations of the main results of our simulations. A theoretical understanding of the origin of the long-range velocity and force correlations in our system is necessary for the development of a theory of the process of development of these correlations. As mentioned earlier, several theoretical studies~\cite{henkes2020,szamel2021,caprini2020a,caprini2020b,kuroda2023} have addressed the mechanism for the occurrence of strong correlations of the velocity field in active systems with long persistence times. These theories are based on a description of the dynamics in terms of the eigenmodes of the Hessian matrix of a local minimum of the potential energy. In the harmonic approximation, the equations of motion of the projections of the displacements of the particles along the eigenvectors of the Hessian matrix reduce, in the overdamped limit, to overdamped Langevin equations in which the noise is correlated over the persistence time of the activity. Solving these equations under the assumption that the low-energy modes of the Hessian matrix are described by elasticity theory, one obtains the result that the length scale of spatial correlations of velocity fluctuations is proportional to the square root of the persistence time $\tau$. This result suggests that this length scale would diverge in the $\tau \to \infty$ limit. This seems to explain the behavior seen in our simulations: an infinite correlation length is expected to show up as a correlation length proportional to the system size in simulations of finite systems. However, a closer examination of the theoretical treatment mentioned above shows that it predicts vanishing velocities for $\tau \to \infty$. This result can be understood from the following argument. If the active forces do not change with time, the dynamics described by Eq.(2) may be thought of as the damped motion of the particles in an effective potential $V_{\rm eff}$ defined as
\begin{equation}
    V_{\rm eff}(\{ {\bf r}_i\}) = \Sigma_{i>j} V_{ij}(|{\bf r}_i - {\bf r}_j|) - \Sigma_i f {\bf r}_i\cdot {\bf n}_i.
\label{veff}    
\end{equation}
The dynamics is then expected converge to a local minimum of $V_{\rm eff}$ (a jammed state) and the velocities are expected to vanish in the long-time limit. Simulations show this behavior if the strength $f$ of the active forces is below a threshold value. However, for larger values of $f$, the system does not converge to jammed state - it is found to fluctuate in the homogeneous liquid state~\cite{mandal2020,liao2018,villarroel2021,yang2022,suman2025} studied here. The reason for this behavior is not clear. A probable explanation is that the effective potential $V_{\rm eff}$, being unbound below for large $\{r_i\}$, does not have any minima when $f$ is large. We are not aware of any theoretical framework that provides a good description of the damped collective dynamics of a system of particles described by a potential function that does nor have any minima. Clearly, such a theoretical framework is necessary for describing the correlations and their development in time found in our simulations.


 The formation of two streams of particles moving in opposite directions in our system may be rationalized by noting that the velocity field in the steady state must satisfy two requirements: the average velocity must be zero and the velocity field must exhibit spatial correlation extending to distances comparable to the system size. Self organization in two counter-propagating streams clearly satisfies these requirements. However, there are many other ways of satisfying these requirements. The formation of two streams moving in opposite directions has the advantage that collisions that alter the spatial arrangement occur only at the interface between the two streams. This structure is, therefore, expected to have a long lifetime. The behavior seen in our system is similar to lane formation is a binary mixture of particles with opposite charges in the presence of a steady electric field~\cite{dzubiella2002}. In this system, for suitable choice of the parameters, particles of the same charge form lanes in which they move in a direction that is parallel or antiparallel to the direction of the electric field, depending on the charge. This is a simpler version of the system we consider, with the ``active'' forces on the charged particles due to the external field pointing in two opposite directions instead of being isotropically distributed. Nevertheless, the theoretical understanding of lane formation~\cite{dzubiella2002,geigenfeind2020} may be helpful in elucidating the behavior seen in our system. The aggregation of particles with similar directions of active forces in our system is analogous to the process of fractionation~\cite{dzubiella2002,fasolo2003} in colloidal suspensions with size polydispersity in which particles with similar sizes come together to form domains in which the particle sizes are nearly the same.  It would be interesting to explore whether a combination of theories of lane formation~\cite{geigenfeind2020} with that of fractionation~\cite{fasolo2003} would lead to a theoretical understanding of the phenomena observed in our simulations.

\section{Acknowledgements}
\label{acknowledgements}
We thank Shiladitya Sengupta for helping us with computational resources. A.S. thanks Manas Chaudhary for discussions which helped in optimizing the simulations. C.D. acknowledges support from the SERB, Department of Science and Technology, India.

\appendix
\label{sec:appendix}

\subsection*{Appendix I: Derivation of Equations \ref{eq:positionupdate} and \ref{eq:velocityupdate}}

We integrate the equation of motion 
\begin{equation*}
    m{\ddot{{\textbf{r}}_i}} = -\gamma{\dot{{\bf{r}}_i}}+\sum_{j\neq i}^{\text{N}}{\textbf{f}}_{ij}+f{{\textbf{n}}_i}
\end{equation*}
over a short interval from $t$ to $t+\Delta t$ in which we assume the force to be constant. Let a temporary variable $s$ represent the time passed \textit{since} $t$ (so $s$ goes from $0$ to $\Delta t$). Over this tiny interval, the velocity update is:
\begin{equation*}
    {\bf{v}}_i(t+s) = {\bf{v}}_i(t)\text{e}^{-\frac{\gamma s}{m}} + \frac{{\bf{f}}_i(t)}{\gamma} \left( 1 - \text{e}^{-\frac{\gamma s}{m}} \right)
\end{equation*}
where ${\bf f}_i(t) = \sum_{j\neq i}^{\text{N}}{\textbf{f}}_{ij}(t)+f{{\textbf{n}}_i}$. Putting $s=\Delta t$ and $\tau = e^{-\frac{\gamma \Delta t}{m}}$, we get Eq.\ref{eq:velocityupdate}. In the first-order approximation, we have $\text{e}^{-\frac{\gamma s}{m}} = 1-\frac{\gamma s}{m}$ and the velocity update simplifies to the forward Euler scheme. Hence, the exponential integrator is more accurate as it retains higher-order terms.

To find the position change, we integrate the velocity over the $\Delta t$ interval:
\begin{equation*}
    {\bf{r}}_i(t+\Delta t) - {\bf{r}}_i(t) = \int_{0}^{\Delta t} {\bf{v}}_i(t+s) \, ds, 
\end{equation*}
\begin{equation*}
    \Delta {\bf{r}}_i = {\bf{v}}_i(t) \int_{0}^{\Delta t} \text{e}^{-\frac{\gamma s}{m}} \, ds + \frac{{\bf{f}}_i(t)}{\gamma} \int_{0}^{\Delta t} \left( 1 - \text{e}^{-\frac{\gamma s}{m}} \right) \, ds
\end{equation*}
Evaluating the first integral,
\begin{equation*}
    \int_{0}^{\Delta t} \text{e}^{-\frac{\gamma s}{m}} \, ds = \frac{m}{\gamma} \left( 1 - \text{e}^{-\frac{\gamma \Delta t}{m}} \right) = \frac{m}{\gamma}(1-\tau) = c_1,
\end{equation*}
and the second integral,
\begin{equation*}
    \int_{0}^{\Delta t} \left( 1 - \text{e}^{-\frac{\gamma s}{m}} \right) \, ds = \Delta t - \frac{m}{\gamma} \left( 1 - \text{e}^{-\frac{\gamma \Delta t}{m}} \right) = \Delta t - \frac{m}{\gamma}(1-\tau),
\end{equation*}
and multiplying the result of the second integral by the $\frac{{\bf f}_i}{\gamma}$ prefactor, we get the term
\begin{equation*}
    \left[ \frac{\Delta t}{\gamma} - \frac{m}{\gamma^2}(1-\tau) \right] {\bf{f}}_i(t).
\end{equation*}
Factoring out $\frac{m}{\gamma^2}$ to match the definition of $c_2$, this term is written as
\begin{equation*}
    \frac{m}{\gamma^2} \left[ \frac{\gamma \Delta t}{m} - 1 + \tau \right] {\bf f}_i(t) = c_2 {\bf f}_i(t).
\end{equation*}
Putting it all together yields the position update equation, Eq.\ref{eq:positionupdate},
\begin{equation}
    {\bf{r}}_i(t+\Delta t) - {\bf{r}}_i(t) = c_1 {\bf{v}}_i(t) + c_2 {\bf{f}}_i(t).
\end{equation}

\subsection*{Appendix II: Comparison of different length scales}
 \begin{figure}[H]
        \centering
        \includegraphics[width=0.79\columnwidth]{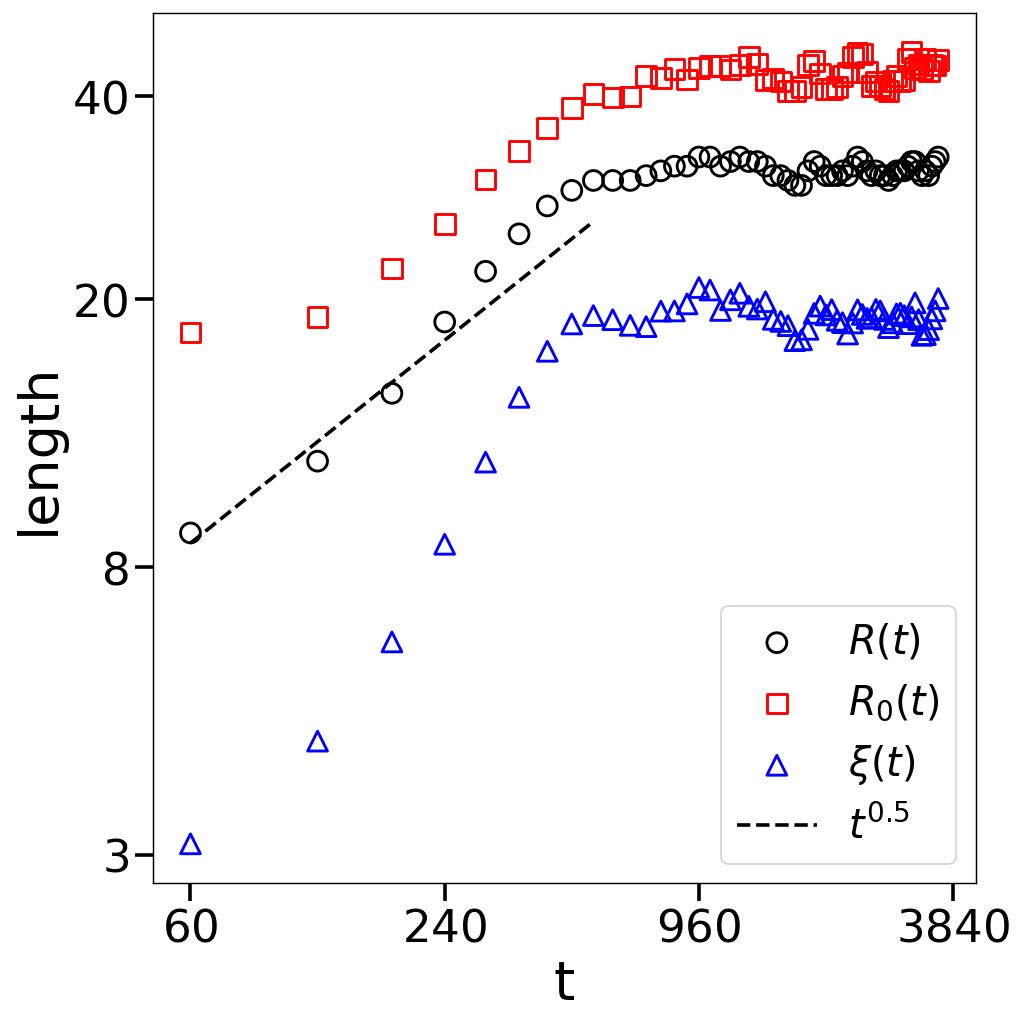}
        \caption{Comparison of correlation lengths for $N=38880$.}
        \label{fig:lengths}
 \end{figure}

\bibliography{references.bib}

@article{janssen2019,
doi = {10.1088/1361-648X/ab3e90},
url = {https://dx.doi.org/10.1088/1361-648X/ab3e90},
year = {2019},
month = {sep},
publisher = {IOP Publishing},
volume = {31},
number = {50},
pages = {503002},
author = {Janssen, Liesbeth M C},
title = {Active glasses},
journal = {Journal of Physics: Condensed Matter}
}

@Inbook{chaudhuri2021,
author="Chaudhuri, Pinaki
and Dasgupta, Chandan",
editor="Meyers, Robert A.",
title="Dense Active Matter",
bookTitle="Encyclopedia of Complexity and Systems Science",
year={2020},
publisher="Springer Berlin Heidelberg",
address="Berlin, Heidelberg",
pages="1--10",
isbn="978-3-642-27737-5",
doi="10.1007/978-3-642-27737-5_713-1",
url="https://doi.org/10.1007/978-3-642-27737-5_713-1"
}

@article{parry2014,
title = {The Bacterial Cytoplasm Has Glass-like Properties and Is Fluidized by Metabolic Activity},
journal = {Cell},
volume = {156},
number = {1},
pages = {183-194},
year = {2014},
issn = {0092-8674},
doi = {https://doi.org/10.1016/j.cell.2013.11.028},
url = {https://www.sciencedirect.com/science/article/pii/S0092867413014797},
author = {Bradley R. Parry and Ivan V. Surovtsev and Matthew T. Cabeen and Corey S. O’Hern and Eric R. Dufresne and Christine Jacobs-Wagner}
}

@Article{nishizawa2017,
author={Nishizawa, Kenji
and Fujiwara, Kei
and Ikenaga, Masahiro
and Nakajo, Nobushige
and Yanagisawa, Miho
and Mizuno, Daisuke},
title={Universal glass-forming behavior of in vitro and living cytoplasm},
journal={Scientific Reports},
year={2017},
month={Nov},
day={09},
volume={7},
number={1},
pages={15143},
issn={2045-2322},
doi={10.1038/s41598-017-14883-y},
url={https://doi.org/10.1038/s41598-017-14883-y}
}

@article{hameed2012,
    doi = {10.1371/journal.pone.0045843},
    author = {Hameed, Feroz M. AND Rao, Madan AND Shivashankar, G. V.},
    journal = {PLOS ONE},
    publisher = {Public Library of Science},
    title = {Dynamics of Passive and Active Particles in the Cell Nucleus},
    year = {2012},
    month = {10},
    volume = {7},
    url = {https://doi.org/10.1371/journal.pone.0045843},
    pages = {1-11},
    number = {10},

}

@article{
angelini2011,
author = {Thomas E. Angelini  and Edouard Hannezo  and Xavier Trepat  and Manuel Marquez  and Jeffrey J. Fredberg  and David A. Weitz },
title = {Glass-like dynamics of collective cell migration},
journal = {Proceedings of the National Academy of Sciences},
volume = {108},
number = {12},
pages = {4714-4719},
year = {2011},
doi = {10.1073/pnas.1010059108},
URL = {https://www.pnas.org/doi/abs/10.1073/pnas.1010059108}}

@article{
garcia2015,
author = {Simon Garcia  and Edouard Hannezo  and Jens Elgeti  and Jean-François Joanny  and Pascal Silberzan  and Nir S. Gov },
title = {Physics of active jamming during collective cellular motion in a monolayer},
journal = {Proceedings of the National Academy of Sciences},
volume = {112},
number = {50},
pages = {15314-15319},
year = {2015},
doi = {10.1073/pnas.1510973112},
URL = {https://www.pnas.org/doi/abs/10.1073/pnas.1510973112}}

@Article{park2015,
author={Park, Jin-Ah
and Kim, Jae Hun
and Bi, Dapeng
and Mitchel, Jennifer A.
and Qazvini, Nader Taheri
and Tantisira, Kelan
and Park, Chan Young
and McGill, Maureen
and Kim, Sae-Hoon
and Gweon, Bomi
and Notbohm, Jacob
and Steward Jr, Robert
and Burger, Stephanie
and Randell, Scott H.
and Kho, Alvin T.
and Tambe, Dhananjay T.
and Hardin, Corey
and Shore, Stephanie A.
and Israel, Elliot
and Weitz, David A.
and Tschumperlin, Daniel J.
and Henske, Elizabeth P.
and Weiss, Scott T.
and Manning, M. Lisa
and Butler, James P.
and Drazen, Jeffrey M.
and Fredberg, Jeffrey J.},
title={Unjamming and cell shape in the asthmatic airway epithelium},
journal={Nature Materials},
year={2015},
month={Oct},
day={01},
volume={14},
number={10},
pages={1040-1048},
issn={1476-4660},
doi={10.1038/nmat4357},
url={https://doi.org/10.1038/nmat4357}
}

@Article{henkes2020,
author={Henkes, Silke
and Kostanjevec, Kaja
and Collinson, J. Martin
and Sknepnek, Rastko
and Bertin, Eric},
title={Dense active matter model of motion patterns in confluent cell monolayers},
journal={Nature Communications},
year={2020},
month={Mar},
day={16},
volume={11},
number={1},
pages={1405},
issn={2041-1723},
doi={10.1038/s41467-020-15164-5},
url={https://doi.org/10.1038/s41467-020-15164-5}
}

@article{lama2024,
    author = {Lama, Hisay and Yamamoto, Masahiro J and Furuta, Yujiro and Shimaya, Takuro and Takeuchi, Kazumasa A},
    title = {Emergence of bacterial glass},
    journal = {PNAS Nexus},
    volume = {3},
    number = {7},
    pages = {pgae238},
    year = {2024},
    month = {06},
    issn = {2752-6542},
    doi = {10.1093/pnasnexus/pgae238},
    url = {https://doi.org/10.1093/pnasnexus/pgae238},
}

@article{klongvessa2019a,
  title = {Active Glass: Ergodicity Breaking Dramatically Affects Response to Self-Propulsion},
  author = {Klongvessa, Natsuda and Ginot, F\'elix and Ybert, Christophe and Cottin-Bizonne, C\'ecile and Leocmach, Mathieu},
  journal = {Phys. Rev. Lett.},
  volume = {123},
  issue = {24},
  pages = {248004},
  numpages = {5},
  year = {2019},
  month = {Dec},
  publisher = {American Physical Society},
  doi = {10.1103/PhysRevLett.123.248004},
  url = {https://link.aps.org/doi/10.1103/PhysRevLett.123.248004}
}

@article{klongvessa2019b,
  title = {Nonmonotonic behavior in dense assemblies of active colloids},
  author = {Klongvessa, Natsuda and Ginot, F\'elix and Ybert, Christophe and Cottin-Bizonne, C\'ecile and Leocmach, Mathieu},
  journal = {Phys. Rev. E},
  volume = {100},
  issue = {6},
  pages = {062603},
  numpages = {9},
  year = {2019},
  month = {Dec},
  publisher = {American Physical Society},
  doi = {10.1103/PhysRevE.100.062603},
  url = {https://link.aps.org/doi/10.1103/PhysRevE.100.062603}
}

@article{arora2022,
  title = {Motile Topological Defects Hinder Dynamical Arrest in Dense Liquids of Active Ellipsoids},
  author = {Arora, Pragya and Sood, A. K. and Ganapathy, Rajesh},
  journal = {Phys. Rev. Lett.},
  volume = {128},
  issue = {17},
  pages = {178002},
  numpages = {6},
  year = {2022},
  month = {Apr},
  publisher = {American Physical Society},
  doi = {10.1103/PhysRevLett.128.178002},
  url = {https://link.aps.org/doi/10.1103/PhysRevLett.128.178002}
}

@article{henkes2011,
  title = {Active jamming: Self-propelled soft particles at high density},
  author = {Henkes, Silke and Fily, Yaouen and Marchetti, M. Cristina},
  journal = {Phys. Rev. E},
  volume = {84},
  issue = {4},
  pages = {040301},
  numpages = {4},
  year = {2011},
  month = {Oct},
  publisher = {American Physical Society},
  doi = {10.1103/PhysRevE.84.040301},
  url = {https://link.aps.org/doi/10.1103/PhysRevE.84.040301}
}

@Article{ni2013,
author={Ni, Ran
and Stuart, Martien A. Cohen
and Dijkstra, Marjolein},
title={Pushing the glass transition towards random close packing using self-propelled hard spheres},
journal={Nature Communications},
year={2013},
month={Oct},
day={28},
volume={4},
number={1},
pages={2704},
issn={2041-1723},
doi={10.1038/ncomms3704},
url={https://doi.org/10.1038/ncomms3704}
}

@article{berthier2014,
  title = {Nonequilibrium Glassy Dynamics of Self-Propelled Hard Disks},
  author = {Berthier, Ludovic},
  journal = {Phys. Rev. Lett.},
  volume = {112},
  issue = {22},
  pages = {220602},
  numpages = {5},
  year = {2014},
  month = {Jun},
  publisher = {American Physical Society},
  doi = {10.1103/PhysRevLett.112.220602},
  url = {https://link.aps.org/doi/10.1103/PhysRevLett.112.220602}
}

@article{berthier2017,
doi = {10.1088/1367-2630/aa914e},
url = {https://dx.doi.org/10.1088/1367-2630/aa914e},
year = {2017},
month = {dec},
publisher = {IOP Publishing},
volume = {19},
number = {12},
pages = {125006},
author = {Berthier, Ludovic and Flenner, Elijah and Szamel, Grzegorz},
title = {How active forces influence nonequilibrium glass transitions},
journal = {New Journal of Physics}
}

@Article{mandal2016,
author ="Mandal, Rituparno and Bhuyan, Pranab Jyoti and Rao, Madan and Dasgupta, Chandan",
title  ="Active fluidization in dense glassy systems",
journal  ="Soft Matter",
year  ="2016",
volume  ="12",
issue  ="29",
pages  ="6268-6276",
publisher  ="The Royal Society of Chemistry",
doi  ="10.1039/C5SM02950C",
url  ="http://dx.doi.org/10.1039/C5SM02950C"}

@article{mandal2017,
  title = {Glassy swirls of active dumbbells},
  author = {Mandal, Rituparno and Bhuyan, Pranab Jyoti and Chaudhuri, Pinaki and Rao, Madan and Dasgupta, Chandan},
  journal = {Phys. Rev. E},
  volume = {96},
  issue = {4},
  pages = {042605},
  numpages = {9},
  year = {2017},
  month = {Oct},
  publisher = {American Physical Society},
  doi = {10.1103/PhysRevE.96.042605},
  url = {https://link.aps.org/doi/10.1103/PhysRevE.96.042605}
}

@article{
nandi2018,
author = {Saroj Kumar Nandi  and Rituparno Mandal  and Pranab Jyoti Bhuyan  and Chandan Dasgupta  and Madan Rao  and Nir S. Gov },
title = {A random first-order transition theory for an active glass},
journal = {Proceedings of the National Academy of Sciences},
volume = {115},
number = {30},
pages = {7688-7693},
year = {2018},
doi = {10.1073/pnas.1721324115},
URL = {https://www.pnas.org/doi/abs/10.1073/pnas.1721324115}}

@article{mandal2022,
doi = {10.1088/2399-6528/ac9c47},
url = {https://dx.doi.org/10.1088/2399-6528/ac9c47},
year = {2022},
month = {nov},
publisher = {IOP Publishing},
volume = {6},
number = {11},
pages = {115001},
author = {Mandal, Rituparno and Nandi, Saroj Kumar and Dasgupta, Chandan and Sollich, Peter and Gov, Nir S},
title = {The random first-order transition theory of active glass in the high-activity regime},
journal = {Journal of Physics Communications}
}

@Article{mandal2020,
author={Mandal, Rituparno
and Bhuyan, Pranab Jyoti
and Chaudhuri, Pinaki
and Dasgupta, Chandan
and Rao, Madan},
title={Extreme active matter at high densities},
journal={Nature Communications},
year={2020},
month={May},
day={22},
volume={11},
number={1},
pages={2581},
issn={2041-1723},
doi={10.1038/s41467-020-16130-x},
url={https://doi.org/10.1038/s41467-020-16130-x}
}

@article{keta2022,
  title = {Disordered Collective Motion in Dense Assemblies of Persistent Particles},
  author = {Keta, Yann-Edwin and Jack, Robert L. and Berthier, Ludovic},
  journal = {Phys. Rev. Lett.},
  volume = {129},
  issue = {4},
  pages = {048002},
  numpages = {7},
  year = {2022},
  month = {Jul},
  publisher = {American Physical Society},
  doi = {10.1103/PhysRevLett.129.048002},
  url = {https://link.aps.org/doi/10.1103/PhysRevLett.129.048002}
}

@Article{keta2023,
author ="Keta, Yann-Edwin and Mandal, Rituparno and Sollich, Peter and Jack, Robert L. and Berthier, Ludovic",
title  ="Intermittent relaxation and avalanches in extremely persistent active matter",
journal  ="Soft Matter",
year  ="2023",
volume  ="19",
issue  ="21",
pages  ="3871-3883",
publisher  ="The Royal Society of Chemistry",
doi  ="10.1039/D3SM00034F",
url  ="http://dx.doi.org/10.1039/D3SM00034F"}

@Article{liao2018,
author ="Liao, Qinyi and Xu, Ning",
title  ="Criticality of the zero-temperature jamming transition probed by self-propelled particles",
journal  ="Soft Matter",
year  ="2018",
volume  ="14",
issue  ="5",
pages  ="853-860",
publisher  ="The Royal Society of Chemistry",
doi  ="10.1039/C7SM01909B",
url  ="http://dx.doi.org/10.1039/C7SM01909B"}

@Article{villarroel2021,
author = "Villarroel, Carlos and During, Gustavo",
title  ="Critical yielding rheology: from externally deformed glasses to active systems",
journal  ="Soft Matter",
year  ="2021",
volume  ="17",
issue  ="43",
pages  ="9944-9949",
publisher  ="The Royal Society of Chemistry",
doi  ="10.1039/D1SM00948F",
url  ="http://dx.doi.org/10.1039/D1SM00948F"}

@article{yang2022,
  title = {Interplay between jamming and motility-induced phase separation in persistent self-propelling particles},
  author = {Yang, Jing and Ni, Ran and Ciamarra, Massimo Pica},
  journal = {Phys. Rev. E},
  volume = {106},
  issue = {1},
  pages = {L012601},
  numpages = {6},
  year = {2022},
  month = {Jul},
  publisher = {American Physical Society},
  doi = {10.1103/PhysRevE.106.L012601},
  url = {https://link.aps.org/doi/10.1103/PhysRevE.106.L012601}
}

@article{suman2025,
      title={Activity-driven sorting, approach to criticality and turbulent flows in dense persistent active fluids}, 
      author={Suman Dutta and Pinaki Chaudhuri and Madan Rao and Chandan Dasgupta},
      year={2025},
      eprint={2509.00376},
      archivePrefix={arXiv},
      primaryClass={cond-mat.soft},
      journal={arXiv preprint},
      url={https://arxiv.org/abs/2509.00376}, 
}

@article{szamel2021,
doi = {10.1209/0295-5075/133/60002},
url = {https://dx.doi.org/10.1209/0295-5075/133/60002},
year = {2021},
month = {may},
publisher = {EDP Sciences, IOP Publishing and Società Italiana di Fisica},
volume = {133},
number = {6},
pages = {60002},
author = {Szamel, Grzegorz and Flenner, Elijah},
title = {Long-ranged velocity correlations in dense systems of self-propelled particles},
journal = {Europhysics Letters}
}

@article{caprini2020a,
  title = {Time-dependent properties of interacting active matter: Dynamical behavior of one-dimensional systems of self-propelled particles},
  author = {Caprini, Lorenzo and Marconi, Umberto Marini Bettolo},
  journal = {Phys. Rev. Res.},
  volume = {2},
  issue = {3},
  pages = {033518},
  numpages = {12},
  year = {2020},
  month = {Sep},
  publisher = {American Physical Society},
  doi = {10.1103/PhysRevResearch.2.033518},
  url = {https://link.aps.org/doi/10.1103/PhysRevResearch.2.033518}
}

@article{caprini2020b,
  title = {Spontaneous Velocity Alignment in Motility-Induced Phase Separation},
  author = {Caprini, L. and Marini Bettolo Marconi, U. and Puglisi, A.},
  journal = {Phys. Rev. Lett.},
  volume = {124},
  issue = {7},
  pages = {078001},
  numpages = {6},
  year = {2020},
  month = {Feb},
  publisher = {American Physical Society},
  doi = {10.1103/PhysRevLett.124.078001},
  url = {https://link.aps.org/doi/10.1103/PhysRevLett.124.078001}
}

@article{kuroda2023,
  title = {Anomalous fluctuations in homogeneous fluid phase of active Brownian particles},
  author = {Kuroda, Yuta and Matsuyama, Hiromichi and Kawasaki, Takeshi and Miyazaki, Kunimasa},
  journal = {Phys. Rev. Res.},
  volume = {5},
  issue = {1},
  pages = {013077},
  numpages = {13},
  year = {2023},
  month = {Feb},
  publisher = {American Physical Society},
  doi = {10.1103/PhysRevResearch.5.013077},
  url = {https://link.aps.org/doi/10.1103/PhysRevResearch.5.013077}
}

@article{bray2002,
author = {A.J. Bray},
title = {Theory of phase-ordering kinetics},
journal = {Advances in Physics},
volume = {43},
number = {3},
pages = {357--459},
year = {1994},
publisher = {Taylor \& Francis},
doi = {10.1080/00018739400101505},
URL = { 
https://doi.org/10.1080/00018739400101505
}
}

@article{bray1989,
  title = {Exact renormalization-group results for domain-growth scaling in spinodal decomposition},
  author = {Bray, A. J.},
  journal = {Phys. Rev. Lett.},
  volume = {62},
  issue = {24},
  pages = {2841--2844},
  numpages = {0},
  year = {1989},
  month = {Jun},
  publisher = {American Physical Society},
  doi = {10.1103/PhysRevLett.62.2841},
  url = {https://link.aps.org/doi/10.1103/PhysRevLett.62.2841}
}

@article{bray1990,
  title = {Renormalization-group approach to domain-growth scaling},
  author = {Bray, A. J.},
  journal = {Phys. Rev. B},
  volume = {41},
  issue = {10},
  pages = {6724--6732},
  numpages = {0},
  year = {1990},
  month = {Apr},
  publisher = {American Physical Society},
  doi = {10.1103/PhysRevB.41.6724},
  url = {https://link.aps.org/doi/10.1103/PhysRevB.41.6724}
}

@article{mondello1993,
  title = {Relaxational dynamics after the quench of a conserved system with a continuous symmetry},
  author = {Mondello, M. and Goldenfeld, Nigel},
  journal = {Phys. Rev. E},
  volume = {47},
  issue = {4},
  pages = {2384--2393},
  numpages = {0},
  year = {1993},
  month = {Apr},
  publisher = {American Physical Society},
  doi = {10.1103/PhysRevE.47.2384},
  url = {https://link.aps.org/doi/10.1103/PhysRevE.47.2384}
}

@article{siegert1993,
  title = {Ordering dynamics of a conserved vector order parameter},
  author = {Siegert, Martin and Rao, Madan},
  journal = {Phys. Rev. Lett.},
  volume = {70},
  issue = {13},
  pages = {1956--1959},
  numpages = {0},
  year = {1993},
  month = {Mar},
  publisher = {American Physical Society},
  doi = {10.1103/PhysRevLett.70.1956},
  url = {https://link.aps.org/doi/10.1103/PhysRevLett.70.1956}
}

@Inbook{porod83,
author="Porod, G",
editor="Glatter, O. and Kratky, O.",
bookTitle="Small angle x-ray scattering",
year={1983},
address="New York",
publisher="Academic Press"
}

@article{dzubiella2002,
  title = {Lane formation in colloidal mixtures driven by an external field},
  author = {Dzubiella, J. and Hoffmann, G. P. and L\"owen, H.},
  journal = {Phys. Rev. E},
  volume = {65},
  issue = {2},
  pages = {021402},
  numpages = {8},
  year = {2002},
  month = {Jan},
  publisher = {American Physical Society},
  doi = {10.1103/PhysRevE.65.021402},
  url = {https://link.aps.org/doi/10.1103/PhysRevE.65.021402}
}

@article{mandal2021,
doi = {10.1088/1361-648X/abef9b},
url = {https://dx.doi.org/10.1088/1361-648X/abef9b},
year = {2021},
month = {apr},
publisher = {IOP Publishing},
volume = {33},
number = {18},
pages = {184001},
author = {Mandal, Rituparno and Sollich, Peter},
title = {How to study a persistent active glassy system},
journal = {Journal of Physics: Condensed Matter}
}

@article{cates2015,
   author = "Cates, Michael E. and Tailleur, Julien",
   title = "Motility-Induced Phase Separation", 
   journal= "Annual Review of Condensed Matter Physics",
   year = "2015",
   volume = "6",
   number = "Volume 6, 2015",
   pages = "219-244",
   doi = "https://doi.org/10.1146/annurev-conmatphys-031214-014710",
   url = "https://www.annualreviews.org/content/journals/10.1146/annurev-conmatphys-031214-014710",
   publisher = "Annual Reviews",
   issn = "1947-5462",
   type = "Journal Article",
  }

@article{fasolo2003,
  title = {Equilibrium Phase Behavior of Polydisperse Hard Spheres},
  author = {Fasolo, Moreno and Sollich, Peter},
  journal = {Phys. Rev. Lett.},
  volume = {91},
  issue = {6},
  pages = {068301},
  numpages = {4},
  year = {2003},
  month = {Aug},
  publisher = {American Physical Society},
  doi = {10.1103/PhysRevLett.91.068301},
  url = {https://link.aps.org/doi/10.1103/PhysRevLett.91.068301}
}

@Article{geigenfeind2020,
author={Geigenfeind, Thomas
and de las Heras, Daniel
and Schmidt, Matthias},
title={Superadiabatic demixing in nonequilibrium colloids},
journal={Communications Physics},
year={2020},
month={Jan},
day={24},
volume={3},
number={1},
pages={23},
issn={2399-3650},
doi={10.1038/s42005-020-0287-5},
url={https://doi.org/10.1038/s42005-020-0287-5}
}

@article{Brüning_2009,
doi = {10.1088/0953-8984/21/3/035117},
url = {https://doi.org/10.1088/0953-8984/21/3/035117},
year = {2008},
month = {dec},
publisher = {},
volume = {21},
number = {3},
pages = {035117},
author = {Brüning, Ralf and St-Onge, Denis A and Patterson, Steve and Kob, Walter},
title = {Glass transitions in one-, two-, three-, and four-dimensional binary Lennard-Jones
systems},
journal = {Journal of Physics: Condensed Matter}
}
\end{document}